\documentclass[12pt]{article}
\usepackage{amsmath,amssymb,cite,epsfig,url,psfrag,eepic,amsmath,mathtools,esint}
\usepackage[bookmarks=true,hyperfigures=true,colorlinks=true,linkcolor=blue,citecolor=blue,urlcolor=blue,bookmarksnumbered]{hyperref}
\usepackage{esint}
\usepackage{bbm}
\usepackage{physics}
\usepackage{amsthm}
\newtheorem{proposition}{Proposition}[section]

\advance \topmargin by -\headheight
\advance \topmargin by -\headsep
\evensidemargin \oddsidemargin
\numberwithin{equation}{section}

\makeatletter
\def\Appendix{\appendix
  \def\@seccntformat##1{Appendix~\csname the##1\endcsname.~~}}
\makeatother

\def\XXint#1#2#3{{\setbox0=\hbox{$#1{#2#3}{\int}$}
\vcenter{\hbox{$#2#3$}}\kern-.5\wd0}}

\usepackage{dynkin-diagrams}

\begin{document}
\title{\textbf{FLZ integrability in the 't Hooft model}\vspace*{0.3cm}}
\date{}
\author{Alexey Litvinov$^{1}$, Egor Shestopalov$^{1,2}$ and Nikita Slavnov$^{3}$
\\[\medskipamount]
\parbox[t]{0.85\textwidth}{\normalsize\it\centerline{1. Skolkovo Institute of Science and Technology, 121205 Moscow, Russia}}
\\
\parbox[t]{0.85\textwidth}{\normalsize\it\centerline{2. Moscow Institute of Physics and Technology, 141700 Dolgoprudny, Russia}}
\\
\parbox[t]{0.85\textwidth}{\normalsize\it\centerline{3. Steklov Mathematical Institute of Russian Academy of Sciences, 119991 Moscow, Russia}}}
\maketitle
\begin{abstract}
    We study the 't Hooft equation describing mesons in large-$N_c$ two-dimensional QCD with arbitrary unequal quark masses. Using its Fredholm formulation and the associated Baxter finite-difference equation, we investigate the relations between spectral determinants and canonical Baxter $Q$-functions. We derive special-point identities for the Baxter solutions by exploiting finite-rank commutator relations of the underlying integral operators. We also determine the spectral residues of the inhomogeneous solutions and obtain exact expressions for eigenfunction normalization factors in terms of values at distinguished complex points. These results establish the unit-residue property of the Baxter expressions entering the Fateev–Lukyanov–Zamolodchikov (FLZ) spectral-determinant identities. As a further application, we prove the nondegeneracy of the meson spectrum for arbitrary unequal quark masses. We further identify a broader class of integral operators exhibiting the same mechanism, which we refer to as FLZ integrability, and illustrate its applicability to the meson spectral problem in Ising field theory. Our approach reveals a connection between the analytic structure of Baxter equations and the theory of completely integrable Fredholm operators.
    \end{abstract}
\tableofcontents
\section{Introduction}
The large-$N_c$ limit of two-dimensional quantum chromodynamics, usually referred to as the 't Hooft model, occupies a special place among analytically tractable models of confinement. Despite the absence of propagating gluonic degrees of freedom in two dimensions, the model retains several characteristic features of four-dimensional QCD: quarks do not appear as asymptotic states, the physical spectrum consists of colorless mesons, and the highly excited states exhibit an approximately linear Regge-type spectrum. At the same time, in the planar limit \cite{tHooft:1973alw}, the bound-state problem reduces to a single integral equation, the celebrated 't Hooft equation \cite{THOOFT1974461}. 

The physical properties of the 't Hooft model were investigated extensively soon after its introduction. In particular, Callan, Coote and Gross \cite{Callan:1976PhysRev} studied the consistency of confinement in the large-$N_c$ limit and developed a description of hadronic scattering and current amplitudes. Einhorn \cite{Einhorn:1976uz} subsequently analyzed electromagnetic form factors and deep-inelastic scattering, demonstrating how sums over the discrete spectrum of confined mesons reproduce the partonic description of high-energy processes. 

For a meson composed of quarks with masses $m_1$ and $m_2$, the spectral problem can be written in the form
\begin{equation}\label{main-eq}
  2\pi^2\lambda\,\varphi(x)
  =
  \left(
     \frac{\alpha_1}{x}
     +\frac{\alpha_2}{1-x}
  \right)\varphi(x)
  -
  \fint_0^1\limits\frac{\varphi(y)\,dy}{(x-y)^2},
  \qquad
  \alpha_i=\frac{\pi m_i^2}{g^2}-1.
\end{equation}
Here $g^2=g_{\mathrm{YM}}^2N_c$ is the 't Hooft coupling, which remains fixed in the planar limit $N_c\to\infty$, with $g_{\mathrm{YM}}$ denoting the Yang–Mills gauge coupling. In two spacetime dimensions, $g$ has the dimension of mass. The parameters $\alpha_i$ and $\lambda$ are therefore dimensionless, and the physical meson masses are related to the eigenvalues by $M_n^2=2\pi g^2\lambda_n$.

The equation \eqref{main-eq} is straightforward to solve numerically to high precision. Its analytic structure, however, is considerably more subtle. Even such apparently elementary questions as the systematic large-$n$ expansion of the spectrum, exact spectral sum rules, and the analytic dependence on the quark masses lead to a surprisingly rich mathematical problem.

An important step toward understanding this structure was made by Fateev, Lukyanov and Zamolodchikov (FLZ) in \cite{Fateev:2009jf}. For the special choice of equal quark masses corresponding to $\alpha_1=\alpha_2=0$ they showed that, after an appropriate Fourier transformation, the 't Hooft equation can be recast into a finite-difference equation of Baxter type. The original spectral problem is then encoded in two distinguished solutions $Q_\pm(u|\lambda)$ of the corresponding $TQ$ equation, selected by their analyticity, asymptotic behavior, and parity properties. This reformulation reveals an integrable structure that is largely hidden in the original $x$-space representation of the problem.

The spectrum can be encoded in the spectral determinants $D_\pm(\lambda)$, whose zeros correspond to the even- and odd-parity sectors, respectively. The expansion of $\log D_\pm(\lambda)$ around $\lambda=0$ generates spectral sums over inverse powers of the meson eigenvalues, with an appropriate regularization of the first sum. Such sums are of considerable physical interest, since they characterize collective properties of the meson spectrum and are closely related to the weighted spectral sums appearing in spectral representations of correlation functions of local operators. In particular, they provide a connection between the discrete bound-state spectrum and the short-distance behavior of gauge-invariant observables, where the predictions of perturbative QCD and the operator product expansion become applicable.

In the special case considered in \cite{Fateev:2009jf}, FLZ proposed the striking identities
\begin{equation}\label{intro:FLZrelation}
 \partial_\lambda\log D_\pm(\lambda)
 =
 2i\,
 \left.
 \partial_u\log Q_\mp(u|\lambda)
 \right|_{u=i}.
\end{equation}
These are Eq.~(2.17) of \cite{Fateev:2009jf}. Their simplicity is rather unexpected. The left-hand side is a genuinely global spectral quantity: through the spectral determinant, it contains information about the entire infinite sequence of meson masses. The right-hand side, by contrast, requires only the logarithmic derivative of one solution of the Baxter equation at the single distinguished point $u=i$.

The rigorous relation between the Fredholm determinant and the solutions of the inhomogeneous integral equation naturally involves integrals over the whole real $u$-axis. The identity \eqref{intro:FLZrelation} asserts that, for the 't Hooft kernel, these non-local expressions collapse to local data at $u=i$. It is precisely this reduction that makes the analytic calculation of high spectral sums and the large-$\lambda$ asymptotics practical. In fact, much of the remarkable efficiency of the FLZ approach relies on this identity. 

Recently, the FLZ construction was extended to arbitrary equal quark masses in \cite{Litvinov:2024riz}. In this case, the Baxter equation survives while its analytic structure acquires a non-trivial dependence on the mass parameter $\alpha$. The simple relation \eqref{intro:FLZrelation} is correspondingly deformed. In the notation of \cite{Litvinov:2024riz}, the resulting formulas are Eqs.~(2.32) and (2.33). 

The situation becomes even more interesting for unequal masses. The most general two-flavor problem was considered in \cite{Artemev:2025cev}. Introducing
\begin{equation}
 \alpha=\frac{\alpha_1+\alpha_2}{2},
 \qquad
 \beta=\frac{\alpha_2-\alpha_1}{2},
 \notag
\end{equation}
one finds that the $u\to-u$ parity symmetry of the equal-mass problem is replaced by a combined transformation $u\to-u$, $\beta\to-\beta$. Consequently, the natural objects entering the spectral relations are no longer simply $Q_\pm(u)$, but their particular linear combinations:
$\mathbf Q_\pm(u)$. Nevertheless, the same phenomenon persists:  equations~(5.18) and (5.21) of \cite{Artemev:2025cev} provide nontrivial unequal-mass generalizations of \eqref{intro:FLZrelation}.  Thus, even after the simple parity decomposition has disappeared, the full spectral determinant continues to be controlled by local Baxter data at the same distinguished point.

Related methods have been applied to confinement problems in Ising field theory \cite{Fonseca:2001dc, Fonseca:2006au,Litvinov:2025geb} and large-$N_c$ scalar QCD$_2$ \cite{Meshcheriakov:2025lsx}, where analogous relations between spectral determinants and Baxter functions have been proposed. The Baxter-equation framework has also been extended to more general two-dimensional gauge theories\cite{Ambrosino:2023dik,Ambrosino:2024prz}.  More recently,  the 't Hooft model on two-dimensional anti-de Sitter and de Sitter spacetimes has been investigated \cite{Betzios:2026btz}. The Ising meson problem provides a particularly close parallel to the present construction. As we demonstrate in Section \ref{FLZ-integrability}, its Fredholm kernel possesses the same finite-rank commutator structure, leading to analogous relations between the Baxter functions and spectral residues. These developments suggest that the FLZ identities may reflect structural properties of a broader class of integrable integral operators.

From a mathematical perspective, these developments are closely related to the theory of completely integrable integral operators. A characteristic feature of this theory is that certain Fredholm kernels admit representations with finite-rank numerators, leading to remarkable identities for their resolvents and spectral determinants. Such structures were systematically developed in the work \cite{Its:1990MPhysB} and subsequently found important applications in the theory of quantum correlation functions, random matrices, and Riemann–Hilbert problems. The integral operators arising from the Fourier representation of the 't Hooft equation possess closely related properties. In particular, their commutators with a distinguished multiplication operator are of finite rank. As we demonstrate below, this property plays an essential role in the derivation of the spectral identities and in the separation of the two spectral sectors.

The main purpose of this paper is to clarify the mathematical origin of the FLZ spectral identities for arbitrary unequal quark masses. Exploiting the finite-rank commutator structure of the underlying integral operators, we derive special-point relations between Baxter solutions and identify the combinations associated with the two spectral sectors. We also prove the nondegeneracy of the meson spectrum and obtain explicit formulas for the residues of the inhomogeneous solutions and the normalization factors of meson eigenfunctions. These results establish the unit-residue property of the proposed FLZ expressions at all physical spectral points. The complete proof of the spectral-determinant identities requires, in addition, control of the remaining entire-function ambiguity.

The paper is organized as follows. In Section \ref{section-2}, we review the Fourier-space formulation of the ’t Hooft equation and its associated Baxter equation. Section \ref{operator-form-int-eq} introduces the inhomogeneous integral equations and the finite-rank commutator identities. In Section \ref{nondeg-section}, we establish the nondegeneracy of the spectrum and construct the distinguished Baxter solutions associated with the two spectral sectors. Section \ref{linear-relations-between-Q} derives exact relations between their values at special points, while Section \ref{Wronskian-section} discusses the quantum Wronskian and spectral determinants. In Section \ref{residue-section}, we calculate the residues of the inhomogeneous solutions at the meson eigenvalues. These results are used in Section \ref{log-relations} to establish the unit-residue property underlying the FLZ identities and to identify the remaining analytic ambiguity. Section \ref{FLZ-integrability} discusses possible extensions to a broader class of integrable integral operators. Technical details of the contour calculations are collected in the appendices.
\section{The homogeneous equation and the \texorpdfstring{$Q$}{Q}-function}\label{section-2}
We pass to Fourier space by introducing
\begin{equation}\label{eq:Fourier}
 \hat{\varphi}(u)=\int_0^1\limits\frac{d x}{2x(1-x)}
       \left(\frac{x}{1-x}\right)^{-i u/2}\varphi(x),\qquad
 \varphi(x)=\int_{-\infty}^\infty \limits\frac{d u}{2\pi}
       \left(\frac{x}{1-x}\right)^{i u/2}\hat{\varphi}(u).
\end{equation}
Then the original 't Hooft equation \eqref{main-eq} has the form
\begin{equation} \label{Fredholm_homogeneous}
    f(u) \hat{\varphi}(u)
    - \frac{2 i \beta}{\pi} \fint_{\mathbb{R}}\limits  dv \frac{1}{2 \sinh{\frac{\pi (u-v)}{2}}}\hat{\varphi}(v) 
    = \lambda \int_{\mathbb{R}}\limits d v \frac{ \pi (u-v)}{2 \sinh{\frac{\pi (u-v)}{2}}} 
    \hat{\varphi}(v)
\end{equation}
This is the homogeneous Fredholm equation (2.11) of \cite{Artemev:2025cev}. We use the
same mass parameters and spectral normalization:
\begin{gather}
 \alpha=\frac{\alpha_1+\alpha_2}{2},\qquad
 \beta=\frac{\alpha_2-\alpha_1}{2},\qquad
 \alpha_j=\frac{\pi m_j^2}{g^2}-1,\qquad M_n^2=2\pi g^2\lambda_n,
       \label{eq:masses}\\
 f(u)=\frac{2\alpha}{\pi}+u\coth\frac{\pi u}{2}.
              \label{eq:parameters}
\end{gather}

For the physical range of quark masses $\alpha_1$, $\alpha_2>-1$, or equivalently $1+\alpha>|\beta|$, the function $f(u)$ is strictly positive and behaves as $f(u)\sim |u|$ for $|u|\to\infty$. The natural Hilbert space for the Fourier-space eigenvalue problem \eqref{Fredholm_homogeneous} is $L^2(\mathbb R,f(u)du)$ i.e. a weighted Hilbert space consisting of measurable functions $\phi(u)$ satisfying
\begin{equation}
\|\phi\|^2_f=
\int_{\mathbb{R}}\limits|\phi(u)|^2f(u)\,du<\infty.
\end{equation}
The singular integrals in \eqref{Fredholm_homogeneous} are understood in the following principal-value sense
\begin{equation}\label{contour_def}
    \fint_{\mathbb{R}}\limits=\frac{1}{2} \left( \;\int_{\mathbb{R}-i \epsilon}\limits+\int_{\mathbb{R}+i \epsilon}\limits \; \right).
\end{equation}

If we make the substitution 
\begin{equation}
    f(u)\hat{\varphi}(u)=\Psi(u|\beta),
\end{equation}
then equation \eqref{Fredholm_homogeneous} takes the form
\begin{equation} \label{Fredholm_homogeneous2}
     \Psi(u|\beta) - \frac{i \beta}{\pi} \fint_{\mathbb{R}}\limits \frac{dv}{f(v)} \frac{1}{\sinh{\frac{\pi (u-v)}{2}}} \Psi(v|\beta) = \frac{\pi\lambda}{2} \int_{\mathbb{R}}\limits \frac{dv}{f(v)} \frac{ u-v}{\sinh{\frac{\pi (u-v)}{2}}} \Psi(v|\beta).
\end{equation}
After this rescaling, we work in the weighted Hilbert space
\begin{equation}\label{weighted-Hilbert-space}
\mathcal H=L^2\left(\mathbb R,\frac{du}{f(u)}\right).
\end{equation}
For $\beta=0$, equation \eqref{Fredholm_homogeneous2} is a Fredholm integral equation with a Hilbert–Schmidt kernel. Indeed, after the unitary transformation from $\mathcal H$ to $L^2(\mathbb R,du)$, the integral operator on its right-hand side has the symmetric kernel
\begin{equation}
K_{\mathrm{sym}}(u,v)=
\frac{u-v}{\sqrt{f(u)f(v)}\sinh\frac{\pi(u-v)}2},
\end{equation}
which is square-integrable on $\mathbb R^2$. Consequently, this operator is compact. For $\beta\neq0$, the additional singular integral operator is bounded but not Hilbert–Schmidt. Nevertheless, since $|\beta|<1+\alpha$ the operator in the left-hand side of \eqref{Fredholm_homogeneous2} is strictly positive and invertible. Thus, by inverting, the full equation can again be reduced to a Fredholm equation with a positive Hilbert–Schmidt kernel (see section \ref{Wronskian-section} for further details). Its nonzero eigenvalues can accumulate only at zero. Therefore, the spectrum consists of discrete positive eigenvalues satisfying $\lambda_n\to\infty$ as $n\to\infty$.

Following Eq.~(2.13) of \cite{Artemev:2025cev}, we define
\begin{equation}\label{Q-definition}
 Q(u)=\sinh\frac{\pi u}{2}\Psi(u).
\end{equation}
Let us recall the analytical properties of the $Q$-function. In the strip $|\text{Im } u|\leq2$, the function $Q$ has subexponential
growth:
\begin{equation}
 Q(u)=O\!\left(e^{\epsilon|\text{Re } u|}\right)
 \quad (|\text{Re } u|\to\infty),\qquad \text{for every }\epsilon>0 .
\end{equation}
Let $u=\pm i u_1^*(\alpha)$ denote the pair of imaginary zeros closest to the origin of the equation 
\begin{equation}
    \frac{2 \alpha}{\pi}+u \coth{\frac{\pi u }{2}}=0.
\end{equation}
At generic mass parameters, the required zeros and the only permitted
poles of the homogeneous $Q$-function in the strip are \cite{Artemev:2025cev}
\begin{equation}
    \begin{aligned}
         \text{required zeros:}\quad&
         0,\ \pm2 i,\ \pm i u_1^*(\alpha),                   \\    
         \text{possible poles:}\quad&
         i u_1^*(\alpha+\beta),\ -i u_1^*(\alpha-\beta). \label{poles-of-Q}
    \end{aligned}
\end{equation}
Additional zeros are allowed, and a permitted pole need not be present. The homogeneous equation also imposes the spectral
quantization conditions
\begin{equation}\label{quantization}
 Q(0)=Q(2 i)=Q(-2i)=0 .
\end{equation}

Analytically continuing \eqref{Fredholm_homogeneous2} to $u\pm2i$ and
combining the three equations cancels the integral terms. The diagonal residues give the Baxter $TQ$ equation
\begin{equation}\label{TQ-equation}
 \left(1+\frac{2\beta}{\pi f(u+2i)}\right)Q(u+2i)
 +\left(1-\frac{2\beta}{\pi f(u-2i)}\right)Q(u-2i)
 -2Q(u)=-\frac{4\pi\lambda}{f(u)}Q(u).
\end{equation}
With the notation
\begin{equation}
 x(u)=\frac2\pi\tanh\frac{\pi u}{2},\qquad
 z(u)=2\pi\lambda\tanh\frac{\pi u}{2},
\end{equation}
this is precisely Eq.~(2.20) of \cite{Artemev:2025cev}.
\section{The inhomogeneous equation and operator form}\label{operator-form-int-eq}
Any normalizable solution to a homogeneous equation necessarily satisfies the $TQ$ equation \eqref{TQ-equation}. However, if we take a $Q$-function that satisfies the $TQ$ equation and relax the quantization condition \eqref{quantization}, then, as shown in \cite{Artemev:2025cev}, the $\Psi$ function corresponding to this $Q$-function now satisfies the inhomogeneous Fredholm equation:
\begin{equation}\label{inhomogeneous-equation}
 \Psi(u)-\frac{i\beta}{\pi}\fint_{\mathbb{R}}\limits
       \frac{\Psi(v)}{\sinh\frac{\pi(u-v)}2}\,\frac{d v}{f(v)}
 -\frac{\pi\lambda}{2}\fint_{\mathbb{R}}\limits
       \frac{(u-v)\Psi(v)}{\sinh\frac{\pi(u-v)}2}\,\frac{d v}{f(v)}
 =a e_+(u)+b e_-(u),
\end{equation}
where
\begin{gather}
 e_+(u)=\frac{u}{\sinh\frac{\pi u}{2}},\qquad
 e_-(u)=\frac{1}{\sinh\frac{\pi u}{2}},                            \label{eq:sources}\\
 a=\frac{Q(2i)-Q(-2i)+\frac{2\beta}{1+\alpha}Q(0)}
          {4i},\qquad
 b=Q(0)
 .\label{eq:sourceboundary}
\end{gather}
The source is therefore determined by the values of $Q$ at the three
quantization points. Setting $a=b=0$ recovers the homogeneous problem.
The basis solutions $\Psi_\pm$ correspond to $(a,b)=(1,0)$ and $(0,1)$,
respectively. We write 
\begin{equation}
    Q_\pm(u|\lambda) = \sinh{\frac{\pi u}{2}} \Psi_\pm(u|\lambda).
\end{equation}
These solutions are normalized as in \cite[Eq.~(2.28)]{Artemev:2025cev}:
\begin{equation}\label{eq:Qnormalization}
    \begin{aligned}
     Q_+(0)=0,\qquad Q_+(2i)-Q_+(-2i)=4i,\\
     Q_-(0)=1,\qquad Q_-(2i)-Q_-(-2i)=-\frac{2\beta}{1+\alpha}.
    \end{aligned}
\end{equation}
Both satisfy the same $TQ$ equation \eqref{TQ-equation}. Their symmetry is
\begin{equation}
 Q_\pm(u|\alpha,\beta)=\mp Q_\pm(-u|\alpha,-\beta).
\end{equation}
As functions of $\lambda$, these solutions are meromorphic, with spectral poles at $\lambda_n$. At unequal masses, the non-bold solutions do not separately select the even and odd spectral sectors. The combinations $\boldsymbol Q_\pm$ that do so will be introduced below.

It will be convenient for us to rewrite this equation in operator form
\begin{equation} \label{homogeneous_operator_form}
    \Psi(u|\beta)-\frac{i \beta}{\pi} (\widehat{S} \Psi)(u|\beta)-\frac{\pi \lambda}{2} (\widehat{K} \Psi)(u|\beta)=a e_+(u)+b e_-(u).
\end{equation}
We define 
\begin{equation}\label{kernel-def}
    (\widehat{S} \Psi)(u|\beta) \overset{\text{def}}{=} \fint_{\mathbb{R}}\limits \frac{dv}{f(v)} \frac{1}{ \sinh{\frac{\pi (u-v)}{2}}} \Psi(v|\beta), \quad (\widehat{K} \Psi)(u|\beta) \overset{\text{def}}{=} \fint_{\mathbb{R}}\limits \frac{dv}{f(v)} \frac{u-v}{ \sinh{\frac{\pi (u-v)}{2}}} \Psi(v|\beta).
\end{equation}

The integral operators appearing in \eqref{kernel-def} are closely related to the class of completely integrable Fredholm operators introduced in \cite{Its:1990MPhysB}. Indeed, after the change of variables $x=e^{\pi u}$, the kernel of $\widehat K$ assumes the characteristic form
\begin{equation}
 \frac{\sum_{j=1}^{2}F_j(x)G_j(y)}{x-y}   
\end{equation}
Such kernels possess remarkable algebraic and analytic properties. In particular, their resolvents preserve the integrable structure, and their Fredholm determinants can be related to solutions of associated Riemann–Hilbert problems. These properties underlie many exact results for quantum correlation functions and random matrix models \cite{Its:1990MPhysB,Tracy:1993ah,b95d26cd-9a9f-3e25-9092-a5b7d6af1e73}. In the present problem, the additional operator $\widehat S$ introduces a singular Cauchy-type contribution requiring a principal-value prescription. Nevertheless, the finite-rank commutator identities derived below demonstrate that essential features of the integrable-kernel structure survive in the full unequal-mass problem.

We also introduce the  operator
\begin{equation} \label{L-def}
    \widehat{L}_{\lambda,\beta} = \widehat{\mathbbm{1}}-\frac{i \beta}{\pi} \widehat{S}-\frac{\pi \lambda}{2} \widehat{K}.
\end{equation}
In the Hilbert space
\begin{equation}
\mathcal H=L^2\left(\mathbb R,\frac{du}{f(u)}\right),
\end{equation}
the real formally symmetric kernel $\widehat K$ is  self-adjoint, whereas the real formally antisymmetric singular kernel $\widehat S$ is skew-adjoint
\begin{equation}
\widehat K^\dagger=\widehat K,\qquad
\widehat S^\dagger=-\widehat S,
\end{equation}
thus for real $(\beta,\lambda)$ the operator $\widehat{L}_{\lambda,\beta}$ is self-adjoint.

The transpose of this operator has the form
\begin{equation}
    \widehat{L}^{\text{T}}_{\lambda,\beta} = \widehat{\mathbbm{1}}+\frac{i \beta}{\pi} \widehat{S}-\frac{\pi \lambda}{2} \widehat{K}=\widehat{L}_{\lambda,-\beta}.
\end{equation}
It follows that if $\Psi(u|\beta)$ is a solution of the equation $\widehat{L}_{\lambda,\beta} \Psi=0$, then $\tilde{\Psi}(u|\beta) = \Psi(-u|\beta)$ is a solution of the equation  $\widehat{L}^{\text{T}}_{\lambda,\beta} \Tilde{\Psi}=\widehat{L}_{\lambda,-\beta} \Tilde{\Psi} =0 $.

We introduce the functions $\Phi_n(u)$ and $\Psi_\pm(u)$ (here we omit their dependence on $\beta$ and $\lambda$, but we assume it to be implied throughout the rest of the text) that satisfy the corresponding equations
\begin{equation}
    \widehat{L}_{\lambda_n,\beta} \Phi_n(u)=0, \quad \widehat{L}_{\lambda,\beta} \Psi_\pm(u)=e_\pm(u),
\end{equation}
\begin{equation}
    \Tilde{\Phi}_n(u)=\Phi_n(-u), \quad \Tilde{\Psi}_\pm(u)=\Psi_\pm(-u).
\end{equation}
In particular, $\Psi_-$ has a pole at zero, whereas $\Psi_+$ is regular:
\begin{equation}\label{singular-source-normalization}
 \Psi_+(u)=O(1),\qquad \Psi_-(u)=\frac{2}{\pi u}+O(1),
 \qquad u\to0,
\end{equation}
For a singular source, the $\widehat{K}$ integral in \eqref{inhomogeneous-equation} also
requires the prescription \eqref{contour_def}, despite the removability of its diagonal singularity.

To express the finite-rank commutator relations in a compact form, we introduce bra–ket notation adapted to the weighted Hilbert space $\mathcal H$ \eqref{weighted-Hilbert-space}. A ket $|h\rangle$ represents a function $h(u)$, while a bra $\langle g|$ denotes the linear functional defined by the bilinear pairing
\begin{equation}
\langle g|h\rangle=\int_{\mathbb R}\frac{g(u)h(u)}{f(u)}\,du.
\end{equation}
We emphasize that this pairing is bilinear rather than sesquilinear and should not be confused with the standard Hilbert-space scalar product
\begin{equation}
(g,h)_{\mathcal H} =\int_{\mathbb R}\frac{g(u)^*h(u)}{f(u)}\,du.
\end{equation}

We use the superscript $T$ for the transpose with respect to this bilinear pairing, while $\dagger$ denotes the Hermitian adjoint in $\mathcal H$. When the functions entering these expressions have poles on the integration contour, the pairing is understood with the principal-value prescription \eqref{contour_def}. 

Now, if we define 
\begin{equation}
    M(u)=\coth{\frac{\pi u}{2}},
\end{equation}
then using the elementary identity 
\begin{equation}
    \frac{M(u)-M(v)}{2 \sinh{\frac{\pi(u-v)}{2}}} = -\frac{1}{2} e_-(u) e_-(v),
\end{equation}
one can obtain commutation relations:
\begin{align}
    [\widehat{M},\widehat{S}]&=- \ket{e_-}\bra{e_-}, 
    \\
    [\widehat{M},\widehat{K}]&=-(\ket{e_+}\bra{e_-}-\ket{e_-}\bra{e_+}) ,  
\end{align}
where $\widehat{M}$ denotes multiplication by $\coth{\frac{\pi u}{2}}$. In particular, 
\begin{equation}\label{M-L-comutator}
    [\widehat{M},\widehat{L}_{\lambda,\beta}]=\frac{i \beta}{\pi} \ket{e_-}\bra{e_-}+\frac{\pi \lambda}{2}(\ket{e_+}\bra{e_-}-\ket{e_-}\bra{e_+}).
\end{equation}
\section{Two spectral sectors and nondegeneracy of the spectrum}\label{nondeg-section}
For equal quark masses, the 't Hooft equation \eqref{Fredholm_homogeneous} is invariant under $u\rightarrow -u$, and its eigenfunctions can be classified according to their parity. Although this symmetry is lost for unequal masses, the spectrum still admits a natural decomposition into two sectors continuously connected to the even and odd states of the equal-mass theory. The purpose of this section is to establish this decomposition directly from the homogeneous Fredholm equation \eqref{Fredholm_homogeneous2} and to prove that its eigenvalues are nondegenerate. 

Let us define the moments
\begin{equation}\label{eq:A-B}
\begin{gathered}
 \mathcal{A}^{(n)}_{\pm}=\int_{\mathbb{R}\pm i\epsilon}\limits\frac{\Phi_n(u)}{f(u)\sinh \frac{\pi u}{2}} d u,
 \quad
 \mathcal{B}_n=\int_{\mathbb{R}}\limits\frac{u \cdot \Phi_n(u)}{f(u)\sinh \frac{\pi u}{2}} d u=\braket{e_+}{\Phi_n},\\
 \mathcal{A}_n=\frac{\mathcal{A}^{(n)}_+ + \mathcal{A}^{(n)}_-}{2}=\braket{e_-}{\Phi_n}.
\end{gathered}
\end{equation}
The residue at zero and the homogeneous equation \eqref{Fredholm_homogeneous2} evaluated at zero give, respectively,
\begin{align}
    \mathcal{A}^{(n)}_{+}-\mathcal{A}^{(n)}_{-}&=-2 \pi i \cdot\underset{u=0}{\operatorname{Res}} \frac{\Phi_n(u)}{f(u)\sinh \frac{\pi u}{2}}=-\frac{2 \pi i}{1+\alpha} \Phi_n(0),\\
    \Phi_n(0)+\frac{i \beta}{2 \pi}& \left(\mathcal{A}^{(n)}_{+}+\mathcal{A}^{(n)}_{-} \right)=\frac{\pi \lambda_n}{2} \mathcal{B}_n.
\end{align}
Eliminating $\Phi_n(0)$ yields
\begin{equation}\label{For_overlaps-I}
    (1+\alpha+\beta)\mathcal{A}^{(n)}_{+}-(1+\alpha-\beta)\mathcal{A}^{(n)}_{-}=- i \pi^2\lambda_n \mathcal{B}_n.
\end{equation}

Now, let us  multiply the homogeneous equation \eqref{Fredholm_homogeneous2} by
\begin{equation}
\frac{\coth \frac{\pi u}{2}\Phi_n(-u)}{f(u)},
\end{equation}
and integrate with \eqref{contour_def} (perform the principal value integral). We obtain (see Appendix \ref{integrals} for more details)
\begin{equation}\label{For_overlaps-II}
    \frac{i \beta}{2 \pi}\mathcal{A}^{(n)}_{+}\mathcal{A}^{(n)}_{-}=\frac{\pi \lambda_n}{4} \mathcal{B}_n \left(\mathcal{A}^{(n)}_{+}+\mathcal{A}^{(n)}_{-} \right).
\end{equation}

We also introduce two additional moments
\begin{equation}\label{a-b-definition}
   a_n=\int_{\mathbb{R}}\limits \frac{\Phi_n(u)}{f(u) \cosh{\frac{\pi u}{2}}}du, \quad b_n=\int_{\mathbb{R}}\limits \frac{u \cdot \Phi_n(u)}{f(u) \cosh{\frac{\pi u}{2}}}du.
\end{equation}
Using the same method, we can obtain
\begin{equation}\label{A--B-homogeneous}
    -\frac{i \beta}{2\pi} a_n^2+\frac{\pi \lambda_n}{2}a_nb_n=0,
\end{equation}
and 
\begin{equation}\label{homogeneous-at-i}
    \frac{\alpha+\beta}{\alpha} \Phi_n(i)=\frac{\beta}{\pi} a_n+\frac{\pi \lambda_n}{2}(a_n+i b_n), \quad \frac{\alpha-\beta}{\alpha} \Phi_n(-i)=-\frac{\beta}{\pi} a_n+\frac{\pi \lambda_n}{2}(a_n-i b_n).
\end{equation}
\subsection{Nondegeneracy of the spectrum}
The relation \eqref{A--B-homogeneous} can be used to prove nondegeneracy of the spectrum. Consider the norm of the eigenfunction
\begin{equation}\label{N-norm}
    \mathcal{N}_n= 8\pi\int_0^1\limits|\varphi_n(x)|^2 dx.
\end{equation}
As shown in Appendix \ref{N-evaluation}, the normalization constant $\mathcal{N}_n$ can be expressed through $a_n$ and $b_n$ as
\begin{equation}\label{N-A-B}
    \mathcal{N}_n=\left(\pi\lambda_n-\frac{2\alpha}{\pi}\right)a_n^2-\pi\lambda_n b_n^2.
\end{equation}
Thus, $(a_n,b_n)$ cannot be simultaneously zero.

Now, suppose that a fixed eigenvalue $\lambda_n$ has an eigenspace $E_{\lambda_n}$ of dimension at least two. We can always choose the eigenfunctions to satisfy\footnote{Indeed, for real mass parameters and real $\lambda$, equation \eqref{Fredholm_homogeneous} is invariant under the antilinear involution
\begin{equation*}
\Phi(u)\overset{\mathcal J}{\rightarrow}\Phi(-u)^*.
\end{equation*}
Since $\mathcal J^2=1$, every eigenspace admits a basis of solutions invariant under $\mathcal J$
\begin{equation*}
\Phi_+(u)=\frac{\Phi(u)+\Phi(-u)^*}{2}\qquad
\Phi_-(u)=\frac{\Phi(u)-\Phi(-u)^*}{2i}.
\end{equation*}}
\begin{equation}\label{reality-condition}
\Phi(-u)=\Phi(u)^*,\qquad u\in\mathbb R.
\end{equation}
Since $f(u)$ and $\cosh\frac{\pi u}{2}$ are real and even, it follows that
\begin{equation}
a_n\in\mathbb R,\qquad b_n\in i\mathbb R.
\end{equation}
Moreover, they are subject to the relation \eqref{A--B-homogeneous}, which can be rewritten as
\begin{equation}\label{A--B-homogeneous-2}
a_n\left(ib_n+\frac{\beta}{\pi^2\lambda_n}a_n\right)=0.
\end{equation}

Let $E_{\lambda_n}^{\mathcal{J}}$ be the real vector space fixed by \eqref{reality-condition}. Consider the linear map
\begin{equation}
E_{\lambda_n}^{\mathcal{J}}
\overset{T}{\longrightarrow}\mathbb R^2,
\qquad
T(\Phi)=(a_n,ib_n).
\end{equation}
From \eqref{N-norm} and \eqref{N-A-B} we conclude that $T$ is injective
\begin{equation}
T(\Phi)=0\quad\Longrightarrow\quad\Phi=0.
\end{equation}
Due to \eqref{A--B-homogeneous-2}, the image of $T$  is contained in
\begin{equation}
\{(a_n,ib_n):a_n=0\}
\;\cup\;
\left\{(a_n,ib_n):b_n=i\frac{\beta}{\pi^2\lambda_n}a_n\right\}.
\end{equation}
A vector subspace contained in the union of two distinct lines must be contained in one of them. Since the antilinear involution $\mathcal J$ defines a real structure on the eigenspace, $\dim_{\mathbb C}E_{\lambda_n}=\dim_{\mathbb R}E_{\lambda_n}^{\mathcal J}$. Hence
\begin{equation}
\dim E_{\lambda_n}\leq1.
\end{equation}
This proves the nondegeneracy of the spectrum.
\subsection{Two spectral sectors}
For $\beta=0$ the homogeneous Fredholm equation \eqref{Fredholm_homogeneous2} naturally admits even/odd eigenfunctions. For $\beta\neq0$ one can use the relation \eqref{A--B-homogeneous} to distinguish two sectors continuously connected to the parity sectors at $\beta=0$. Indeed, equation \eqref{A--B-homogeneous} has two solutions:
\begin{equation}
a_{n}=0 \quad \text{or} \quad a_{n}= -\frac{i \pi^2 \lambda_n}{\beta} b_{n}.  
\end{equation}
Due to \eqref{N-A-B} both $a_n$ and $b_n$ cannot vanish simultaneously.   By continuity in $\beta$, we have $a_{2n+1}=0$ and $a_{2n}\neq 0$. Thus
\begin{equation}
      a_{2n+1}=0 \quad \text{and} \quad a_{2n}= -\frac{i \pi^2 \lambda_n}{\beta} b_{2n}.  
\end{equation}

For the moments $\mathcal{A}^{(n)}_{\pm}$ and $\mathcal{B}_n$, equations \eqref{For_overlaps-I} and \eqref{For_overlaps-II} imply
\begin{equation}
  (1+\alpha+\beta)\left(\mathcal{A}_+^{(n)}\right)^2=(1+\alpha-\beta)\left(\mathcal{A}_-^{(n)}\right)^2
  \implies
  \left(\mathcal{A}_+^{(n)}\right)^2=
\frac{\gamma^2}{(1+\alpha+\beta)^2}
  \left(\mathcal{A}_-^{(n)}\right)^2,
\end{equation}
where we use the parameter
\begin{equation}
     \gamma=\sqrt{(1+\alpha)^2-\beta^2}.
\end{equation}
For equal masses ($\beta=0$), reflection of the integrand in \eqref{eq:A-B} gives
$\mathcal A^{(n)}_+=(-1)^{n+1}\mathcal A^{(n)}_-$.
Continuation from $\beta=0$ fixes
\begin{equation}\label{homogeneous-branch}
 \mathcal A^{(n)}_+
 =(-1)^{n+1}\frac{\gamma}{1+\alpha+\beta}\mathcal A^{(n)}_-.
\end{equation}
We note that \eqref{homogeneous-branch} coincides with the parity symmetry in \cite{Callan:1976PhysRev}. In $x-$space it reads
\begin{equation}\label{phi-nontrivial-sym}
    \sqrt{1+\alpha-\beta} \int \limits_0^1 \frac{\varphi^{(n)}(x|\beta)}{x} dx = (-1)^n\, \sqrt{1+\alpha+\beta}  \int \limits_0^1 \frac{\varphi^{(n)}(x|\beta)}{1-x} dx.
\end{equation}

Because $f(u)$ is even and  the kernels in \eqref{kernel-def} are respectively odd and even under simultaneous reflection of $u$ and $v$, we find that if $\Phi_n(u|\beta)$ is an eigenfunction with eigenvalue $\lambda_n(\beta)$, then $\Phi_n(-u|\beta)$ is an eigenfunction with $-\beta$ at the same eigenvalue. Due to nondegeneracy, we have
\begin{equation}
\Phi_n(u|\beta)=c_n(\beta)\Phi_n(-u|-\beta)
\end{equation}
for some nonzero scalar $c_n(\beta)$. Choose the corresponding $x$-space eigenfunctions $\varphi^{(n)}(x|\beta)$ to be real, normalized, and continuous in $\beta$. Then \eqref{phi-nontrivial-sym} implies $c_n(\beta)=\pm 1$.  Thus
\begin{equation}
 \Phi_n(u|\beta)=(-1)^n\Phi_n(-u|-\beta),\qquad
 \lambda_n(\beta)=\lambda_n(-\beta).
\end{equation}

We note that the relation \eqref{homogeneous-branch} can be rewritten as
\begin{equation}\label{overlaps}
    \frac{\braket{\Phi_n}{e_+}}{\braket{\Phi_n}{e_-}}=\frac{2 i}{\pi^2 \beta \lambda_n}\left(\beta^2-(1+\alpha)^2 +(-1)^{n+1} (1+\alpha) \gamma \right).
\end{equation}
More specifically, in the even and odd sectors, respectively, we have
\begin{equation}\label{overlaps-even-and-odd}
    \frac{\braket{\Phi_{2n}}{e_+}}{\braket{\Phi_{2n}}{e_-}}= -\frac{2i\gamma(1+\alpha+\gamma)}{\pi^2\beta\lambda_{2n}}, \quad \frac{\braket{\Phi_{2n+1}}{e_+}}{\braket{\Phi_{2n+1}}{e_-}}=\frac{2i\beta\gamma}{\pi^2\lambda_{2n+1}(1+\alpha+\gamma)}.
\end{equation}
\subsection{Bold \texorpdfstring{$\Psi$}{Psi} functions}
Using \eqref{M-L-comutator} let us apply the operator $[\widehat{M},\widehat{L}_{\lambda_n,\beta}]$ to $\ket{\Phi_n}$ which satisfies the equation $\widehat{L}_{\lambda_n,\beta} \ket{\Phi_n}=0$. We obtain 
\begin{equation}\label{comutator_relation}
    \widehat{L}_{\lambda_n,\beta} \left(M \Phi_n \right)(u)=-\frac{\pi \lambda_n}{2} \braket{\Phi_n}{e_-} e_+(u)+\left( \frac{\pi \lambda_n}{2} \braket{\Phi_n}{e_+} -\frac{i \beta}{\pi} \braket{\Phi_n}{e_-}\right)e_-(u).
\end{equation}
Then using \eqref{overlaps-even-and-odd} one can rewrite this relation for both sectors:
\begin{align}
    \widehat{L}_{\lambda_{2n+1},\beta} \left(M \Phi_{2n+1} \right)(u)&=-\frac{\pi \lambda_{2n+1}}{2} \braket{\Phi_{2n+1}}{e_-} \mathbf{e}_+(u|\lambda_{2n+1}), 
                           \label{e_+_bold}
    \\ \widehat{L}_{\lambda_{2n},\beta} \left(M \Phi_{2n} \right)(u)&=\frac{\pi \lambda_{2n}}{2} \frac{1+\alpha}{\gamma} \braket{\Phi_{2n}}{e_+} \mathbf{e}_-(u|\lambda_{2n}),
                       \label{e_-_bold}
\end{align}
where 
\begin{equation}
    \mathbf{e}_+(u|\lambda) \overset{\text{def}}{=} e_+(u)+\frac{2 i \beta}{\pi^2 \lambda} \frac{1+\alpha}{1+\alpha+\gamma} e_-(u), \quad  \mathbf{e}_-(u|\lambda) \overset{\text{def}}{=} e_-(u)-\frac{i \pi^2 \beta \lambda}{2 (1+\alpha)(1+\alpha+\gamma)}  e_+(u).
\end{equation}
Note that the coefficients in these linear combinations coincide with the coefficients in the definition of bold $Q$-functions in (5.13) of \cite{Artemev:2025cev}. Accordingly, we define
\begin{equation}\label{bold-Psi-def}
\begin{aligned}
    &\mathbf{\Psi}_+(u) \overset{\text{def}} {=}\widehat{L}^{-1}_{\lambda,\beta} \mathbf{e}_+= \Psi_+(u)+\frac{2 i \beta}{\pi^2 \lambda} \frac{1+\alpha}{1+\alpha+\gamma} \Psi_-(u),\\  
    &\mathbf{\Psi}_-(u) \overset{\text{def}}{=}\widehat{L}^{-1}_{\lambda,\beta} \mathbf{e}_-= \Psi_-(u)-\frac{i \pi^2 \beta \lambda}{2 (1+\alpha)(1+\alpha+\gamma)}  \Psi_+(u).
\end{aligned}    
\end{equation}
Following the approach in the article \cite{Artemev:2025cev}, we define the bold $Q$-functions as
\begin{equation}\label{bolds-Qs}
    \mathbf{Q}_\pm(u)=\sinh{\frac{\pi u}{2}} \mathbf{\Psi}_\pm(u).
\end{equation}
These functions are analogs of the $Q_\pm$-functions for $\beta=0$ from \cite{Litvinov:2024riz}.  At the physical eigenvalues, the distinguished solutions are regular in one sector and have simple spectral poles in the other. Their special-point values at $u=\pm i$ exhibit complementary zeros, as shown in Section \ref{Poles and zeros of the bold Q-functions}.

Throughout the remainder of the paper, $Q_\pm(u)$ denote the canonical solutions normalized by \eqref{eq:Qnormalization}, whereas $\mathbf Q_\pm(u)$ denote the distinguished linear combinations \eqref{bolds-Qs}. The latter are adapted to the two spectral sectors. This distinction is essential: the canonical solutions generally have poles at both types of eigenvalues, while the distinguished combinations separate their spectral singularities.
\section{Linear relations between \texorpdfstring{$\mathbf{Q}_\pm(i)$}{Q(i)} and \texorpdfstring{$\mathbf{Q}_\pm(-i)$}{Q(-i)}}\label{linear-relations-between-Q}
In this section, we derive the two special-point ratios directly from the inhomogeneous
equations \eqref{inhomogeneous-equation}. Written out using
\eqref{kernel-def}, these equations are
\begin{align}
 \Psi_+(u)&-\frac{i\beta}{\pi}\fint_{\mathbb{R}}\limits
 \frac{\Psi_+(v)}{f(v)\sinh\frac{\pi(u-v)}2}dv
 -\frac{\pi\lambda}{2}\fint_{\mathbb{R}}\limits
 \frac{(u-v)\Psi_+(v)}{f(v)\sinh\frac{\pi(u-v)}2}dv
 =\frac{u}{\sinh\frac{\pi u}2},\label{ratio-inhomogeneous-plus}\\
 \Psi_-(u)&-\frac{i\beta}{\pi}\fint_{\mathbb{R}}\limits
 \frac{\Psi_-(v)}{f(v)\sinh\frac{\pi(u-v)}2}dv
 -\frac{\pi\lambda}{2}\fint_{\mathbb{R}}\limits
 \frac{(u-v)\Psi_-(v)}{f(v)\sinh\frac{\pi(u-v)}2}dv
 =\frac{1}{\sinh\frac{\pi u}2}.\label{ratio-inhomogeneous-minus}
\end{align}

\begin{proposition}\label{linear-relations-prop}
    There exist two simple relations between bold $Q$-functions at points $u=\pm i$:
    \begin{align}\label{linear-relations-for-bold-Q}
    \frac{\mathbf{Q}_-(i)}{\mathbf{Q}_-(-i)}&=\frac{\alpha-\beta}{\alpha+\beta},\\\label{linear-relations-for-bold-Q-2}
    \frac{\mathbf{Q}_+(i)}{\mathbf{Q}_+(-i)}&=\frac{\beta+\pi^2 \lambda}{\beta-\pi^2 \lambda}\,\,\frac{\alpha-\beta}{\alpha+\beta}.
\end{align}
\end{proposition}
\begin{proof}
    We first obtain one quadratic identity from each equation \eqref{ratio-inhomogeneous-plus} and \eqref{ratio-inhomogeneous-minus}, and then a
third identity by cross-multiplication and addition. Evaluation of the
same equations at $u=i$ and $u=-i$ will express their moments through
the special-point values of $Q_+$ and $Q_-$. All explicit computations used in the proof are presented in Appendix \ref{details}.
The subscript $\pm$ labels the solution $\Psi_\pm$ in both families. 

Introduce the moments
\begin{equation}\label{A-B-inhomogeneous}
 A_\pm=\fint_{\mathbb{R}}\limits\frac{\Psi_\pm(u)}{f(u)\cosh\frac{\pi u}2}\,du,
 \qquad
 B_\pm=\fint_{\mathbb{R}}\limits\frac{u \cdot \Psi_\pm(u)}{f(u)\cosh\frac{\pi u}2}\,du.
\end{equation}
These are the overlaps with the sources in
\eqref{ratio-inhomogeneous-plus}--\eqref{ratio-inhomogeneous-minus} after multiplication by $\tanh \frac{\pi u}{2}$. 

First, we multiply \eqref{ratio-inhomogeneous-plus}-\eqref{ratio-inhomogeneous-minus} by
$\tanh\frac{\pi u}{2}\Psi_+(-u)/f(u)$ and by $\tanh\frac{\pi u}{2}\Psi_-(-u)/f(u)$, respectively, and take the contour half-sum. This allows us to express $A_\pm$ and $B_\pm$ in terms of each other:
\begin{align}
    &-\frac{i\beta}{2\pi}A_+^2
 +\frac{\pi\lambda}{2}A_+B_+=-B_+,\label{A-B-plus}\\
  &-\frac{i\beta}{2\pi}
 \left(A_-^2-\frac{\pi^2}{(1+\alpha)^2}\right)
 +\frac{\pi\lambda}{2}A_-B_-=A_-. \label{A-B-minus}
\end{align}

Then we multiply the equation with the source $e_+(u)$ by
$\tanh \frac{\pi u}{2}\Psi_-(-u)/f(u)$, and the equation with the source $e_-$
by $\tanh\frac{\pi u}{2}\Psi_+(-u)/f(u)$. Add the two equations before integrating. One obtains another independent equality 
\begin{equation}\label{A-B-mixed}
     -\frac{i\beta}{\pi}A_+A_-
 +\frac{\pi\lambda}{2}(A_+B_-+A_-B_+)=A_+-B_-.
\end{equation}

We next use the same inhomogeneous equations \eqref{inhomogeneous-equation} at complex $u$.
The source pole in the $v$ integration continues to be averaged as in \eqref{contour_def}. Substituting $u=\pm i$ into the inhomogeneous equations (with $(a,b)=(1,0)$ and $(a,b)=(0,1)$) requires a proper meromorphic extension of the function $(\widehat{S} \Psi_\pm)(u)$ beyond the real axis. This extension has the form
\begin{align}
    (\widehat{S} \Psi_\pm)(i) &= \frac{1}{i} \fint_{\mathbb{R}}\limits \frac{\Psi_\pm (v)}{f(v) \cosh{\frac{\pi v}{2}}}dv + 2i \frac{\Psi_\pm(i)}{f(i)} =-i A_\pm +\frac{i \pi}{\alpha} \Psi_\pm(i),\\
    (\widehat{S} \Psi_\pm)(-i) &= \frac{-1}{i} \fint_{\mathbb{R}}\limits \frac{\Psi_\pm (v)}{f(v) \cosh{\frac{\pi v}{2}}}dv - 2i \frac{\Psi_\pm(-i)}{f(-i)}=i A_\pm -\frac{i \pi}{\alpha} \Psi_\pm(-i).
\end{align}
The signs agree with \eqref{T-inner-contour-transfer}. No diagonal
correction occurs for $\widehat K$:
\begin{equation}
    (\widehat K\Psi_\pm)(i)=A_\pm+iB_\pm, \quad(\widehat K\Psi_\pm)(-i)=A_\pm-iB_\pm.
\end{equation}
Carrying out this substitution in all four cases gives
\begin{equation}\label{inhomogeneous-at-i}
 \begin{aligned}
 \frac{\alpha+\beta}{\alpha}\Psi_+(i)
 &=1+\frac{\beta}{\pi}A_++\frac{\pi\lambda}{2}(A_++iB_+), &\quad 
 \frac{\alpha-\beta}{\alpha}\Psi_+(-i)
 &=1-\frac{\beta}{\pi}A_++\frac{\pi\lambda}{2}(A_+-iB_+),\\
 \frac{\alpha+\beta}{\alpha}\Psi_-(i)
 &=-i+\frac{\beta}{\pi}A_-+\frac{\pi\lambda}{2}(A_-+iB_-), &\quad 
 \frac{\alpha-\beta}{\alpha}\Psi_-(-i)
 &=i-\frac{\beta}{\pi}A_-+\frac{\pi\lambda}{2}(A_--iB_-).
 \end{aligned}
\end{equation}

Using $Q_\pm(i)=i\Psi_\pm(i)$ and
$Q_\pm(-i)=-i\Psi_\pm(-i)$,  subtract each
pair in \eqref{inhomogeneous-at-i}. This gives
\begin{equation}\label{ratio-individual-boundary-moments}
 \begin{aligned}
 (\alpha+\beta)Q_+(i)-(\alpha-\beta)Q_+(-i)
 &=i\alpha(2+\pi\lambda A_+),\\
 (\alpha+\beta)Q_-(i)-(\alpha-\beta)Q_-(-i)
 &=i\alpha\pi\lambda A_-.
 \end{aligned}
\end{equation}
Let us consider the ratio 
\begin{equation}
    \frac{(\alpha+\beta)Q_-(i)-(\alpha-\beta)Q_-(-i)}{(\alpha+\beta)Q_+(i)-(\alpha-\beta)Q_+(-i)}=\frac{\pi \lambda A_-}{2+\pi \lambda A_+}.
\end{equation}
Using identities \eqref{A-B-plus}, \eqref{A-B-minus}, and
\eqref{A-B-mixed} one can express $A_+$ and $B_\pm$ in terms of $A_-$ and substitute them into the right-hand side of the relation above. Then we get
\begin{equation} \label{linear-Q-rel-1}
     \frac{(\alpha+\beta)Q_-(i)-(\alpha-\beta)Q_-(-i)}{(\alpha+\beta)Q_+(i)-(\alpha-\beta)Q_+(-i)}=\frac{i \beta \pi^2 \lambda}{2(1+\alpha)(1+\alpha+\gamma)},
\end{equation}
which is precisely the  expression (5.10) from \cite{Artemev:2025cev}. The second linear relation (5.11) involving the $Q$-function from the same article is obtained in exactly the same way:
\begin{equation}\label{linear-Q-rel-2}
    \left(1-\frac{\pi^2 \lambda}{\beta} \right) Q_+(i)-2i \frac{\alpha-\beta}{\alpha+\beta}\frac{1+\alpha}{\gamma} Q_-(-i)-\frac{\alpha-\beta}{\alpha+\beta} \left(1+\frac{1+\alpha}{\gamma}\frac{\pi^2 \lambda}{\beta} \right) Q_+(-i)=0.
\end{equation}

In terms of $\mathbf{Q}_\pm(\pm i)$ the relations \eqref{linear-Q-rel-1} and \eqref{linear-Q-rel-2} can be rewritten in the form stated in 
Proposition \ref{linear-relations-prop}.
\end{proof}
\section{The quantum Wronskian and spectral determinants}\label{Wronskian-section}
In this section, we relate the analytic properties of the Baxter $Q$-functions to the spectral structure of the ’t Hooft equation. We first consider the quantum Wronskian of the two distinguished solutions and use its factorization to obtain an exact relation between their values at $u=i$. We then introduce the spectral determinants associated with the two spectral sectors. These results will play a central role in the subsequent analysis of the FLZ identities.
\subsection{The quantum Wronskian}
Following (5.4) of \cite{Artemev:2025cev} we define the quantum Wronskian
\begin{equation}\label{Wronskian-definition}
 W(u)=Q_+(u+i)Q_-(u-i)-Q_-(u+i)Q_+(u-i).
\end{equation}
Eliminating the central values with \eqref{TQ-equation} gives
\begin{equation}\label{Wronskian-difference}
 \frac{W(u+i)}{W(u-i)}
 =\frac{u-2i+(\alpha-\beta)x(u)}{u-2i+\alpha x(u)}
  \frac{u+2i+\alpha x(u)}{u+2i+(\alpha+\beta)x(u)},
 \qquad x(u)=\frac2\pi \tanh{\frac{\pi u}{2}}.
 \end{equation}
 The normalized  factorization with the stated analytic
conditions is Eq.~(5.6) of \cite{Artemev:2025cev}:
\begin{equation}\label{Wronskian-full}
 W(u+i)=\frac{2i\gamma}{1+\alpha}
 \frac{u+\alpha x(u)}{u+(\alpha+\beta)x(u)}
 \frac{u+2i+\alpha x(u)}{u+2i+(\alpha+\beta)x(u)}G(u).
\end{equation}
Here $G(u)=S_0(u|\alpha+\beta)/S_0(u|\alpha-\beta)$ is the 
function of Eq.~(3.9) from that paper, where $S_0(u)$ is given by the integral
\begin{equation}\label{S0-integral-representation}
    S_0(u+i|\alpha)=
    \exp\left[\frac{i}{4}\int_{\mathbb{R}}\limits\log
    \left(\frac{4\pi\tanh\left(\frac{\pi t}{2}\right)}{t+\frac{2\alpha}{\pi}\tanh\left(\frac{\pi t}{2}\right)}\right)
    \left(\tanh\frac{\pi(t-u)}{2}-
    \tanh\frac{\pi t}{2}\right)dt\right]
\end{equation}
It satisfies
\begin{equation}
 G(i)=1,\qquad
 G(u)G(-u)=\frac{u+(\alpha+\beta)x(u)}{u+(\alpha-\beta)x(u)},
 \qquad G(-i)=\frac{\alpha+\beta}{\alpha-\beta}.
\end{equation}
The normalization $W(i)=Q_+(2i)Q_-(0)=2i$ follows from
\eqref{eq:Qnormalization} and \eqref{TQ-equation}.
Taking $u=-i$ in \eqref{Wronskian-full},
\begin{equation}\label{Wronskian-at-zero}
 W(0)=Q_+(i)Q_-(-i)-Q_+(-i)Q_-(i)=\frac{2i\gamma}{1+\alpha}\frac{\alpha^2}{\alpha^2-\beta^2}.
\end{equation}
We use this  Wronskian result from the original paper. Using \eqref{linear-Q-rel-1} and \eqref{linear-Q-rel-2} one can express everything in terms of $Q_\pm(i)$ and obtain a factorized form of $W(0)$:
\begin{equation}\label{Wronskian-at-zero-2}
     W(0)=P(\lambda)\mathbf Q_+(i|\lambda)\mathbf Q_-(i|\lambda), \quad P(\lambda)=\frac{\pi^2\lambda}{\pi^2\lambda+\beta}\,
 \frac{\alpha+\beta}{\alpha-\beta}\,
 \frac{1+\alpha+\gamma}{\gamma}.
\end{equation}
\subsection{The spectral determinants}
We now introduce the spectral determinants. The homogeneous spectral problem \eqref{Fredholm_homogeneous2} is equivalent to
\begin{equation}
(1-\lambda \widehat{B}_\beta)\chi=0,
\qquad
\widehat{B}_\beta=\frac{\pi}{2}\widehat{A}_\beta^{-1/2}
\widehat K \widehat{A}_\beta^{-1/2},
\qquad
\widehat{A}_\beta=1-\frac{i\beta}{\pi}\widehat S,
\qquad \chi=A_\beta^{1/2}\Psi.
\end{equation}
For physical mass parameters $\alpha_1,\alpha_2>-1$, the operator $B_\beta$ is positive, self-adjoint, and Hilbert–Schmidt.

Since a Hilbert–Schmidt operator need not be of trace class, its ordinary Fredholm determinant is not necessarily defined. Instead, one introduces the regularized Fredholm determinant
\begin{equation}
D(\lambda)
=\det\bigl((1-\lambda \widehat{B}_\beta)e^{\lambda \widehat{B}_\beta}\bigr).
\end{equation}
Equivalently,
\begin{equation}\label{Fredholm-D}
D(\lambda)
=\prod_{n=0}^{\infty}
\left(1-\frac{\lambda}{\lambda_n}\right)
e^{\frac{\lambda}{\lambda_n}}.
\end{equation}
The infinite product in \eqref{Fredholm-D} converges uniformly on compact subsets of the complex $\lambda$-plane because $\sum_n\lambda_n^{-2}<\infty$. In particular, $D(\lambda)$ is an entire function whose zeros coincide with the physical meson spectrum.

It is convenient to describe the spectrum through the spectral determinants $D_\pm(\lambda)$, which retain information about the two spectral branches continuously connected to the even and odd sectors at $\beta=0$. They were defined in \cite{Litvinov:2024riz} as 
\begin{equation}\label{D_pm-def}
    D_+(\lambda)  \overset{\text{def}}{=} \left(\frac{8 \pi}{e} \right)^\lambda \prod_{n=0}^\infty \left(1-\frac{\lambda}{\lambda_{2n}} \right) e^{\frac{\lambda}{n+1}}, \quad D_-(\lambda)  \overset{\text{def}}{=} \left(\frac{8 \pi}{e} \right)^\lambda \prod_{n=0}^\infty \left(1-\frac{\lambda}{\lambda_{2n+1}} \right) e^{\frac{\lambda}{n+1}}.
\end{equation}
This definition can be written in terms of spectral sums:
\begin{equation}
    D_\pm(\lambda)=\left(\frac{8 \pi}{e} \right)^\lambda  \exp \Biggl[ -\sum_{s=1}^\infty  s^{-1} G_\pm^{(s)} \lambda^s \Biggl],
\end{equation}
where 
\begin{equation}
    G_+^{(s)}=\sum_{n=0}^\infty \left(\frac{1}{\lambda_{2n}^s}-\frac{\delta_{s,1}}{n+1} \right), \quad  G_-^{(s)}=\sum_{n=0}^\infty \left(\frac{1}{\lambda_{2n+1}^s}-\frac{\delta_{s,1}}{n+1} \right).
\end{equation}

Exact analytical expressions for first spectral sums $G_{\pm}^{(s)}$ can be found in \cite{Artemev:2025cev,Litvinov:2024riz,Fateev:2009jf}, as well as for Ising field theory in \cite{Litvinov:2025geb} and scalar QCD$_2$ in \cite{Meshcheriakov:2025lsx} . One nontrivial application of these spectral sums is the study
of critical points under analytic continuation in the mass parameters. In the equal-mass 't Hooft model, their singularities predict an infinite family of critical points on the second
sheet of the complex $\alpha$-plane, at each of which three mesons---one even and two odd---become massless \cite{Litvinov:2025geb}. Related critical structures arise in the two-particle approximation to Ising field theory and in large-$N_c$ scalar QCD$_2$. The physical interpretation of these
families, including their possible relation to nonunitary conformal field theories, remains an interesting open problem.

The apparent difference between the two definitions \eqref{Fredholm-D} and \eqref{D_pm-def} is the replacement
\begin{equation}\label{D-replacement}
    e^{\frac{\lambda}{\lambda_{2n}}}\rightarrow
    e^{\frac{\lambda}{n+1}},\quad
    e^{\frac{\lambda}{\lambda_{2n+1}}}\rightarrow
    e^{\frac{\lambda}{n+1}}
\end{equation}
and an overall factor $\left(\frac{8 \pi}{e} \right)^\lambda$. The large $n$ expansion of $\lambda_n$ has been derived in \cite{Artemev:2025cev}
\begin{equation}
\lambda_n=
\frac12\left(n+\frac34\right)
+\frac{\alpha_1+\alpha_2}{\pi^2}\log n
+\frac{C(\alpha_1)+C(\alpha_2)}{\pi^2}
+O\left(\frac{\log n}{n}\right).
\end{equation}
Therefore,
\begin{equation}
\frac1{\lambda_{2n+\delta}}-\frac1{n+1}
= O\left(\frac{\log n}{n^2}\right),\quad
\delta=0,1.
\end{equation}
This estimate guarantees the convergence of the regularized products in \eqref{D_pm-def}.
\section{Spectral residues of the inhomogeneous solutions}\label{residue-section}
In this section, we study the behavior of the solutions of the inhomogeneous integral equation \eqref{inhomogeneous-equation} near the eigenvalues of the ’t Hooft Hamiltonian. At each spectral point, the corresponding solution develops a simple pole whose residue is proportional to the associated eigenfunction. Namely, we have the following proposition:
\begin{proposition}
    The functions $\Psi_\pm(u|\lambda)$ have poles at the spectral points $\lambda=\lambda_n$, with residues given by 
          \begin{align}
        \underset{\lambda=\lambda_n}{\operatorname{Res}} \Psi_+(u)&=-\frac{2 \Phi_n(u)}{\pi} \frac{\braket{\Phi_n}{e_+}}{\langle\tilde{\Phi}_n |\widehat{K} |\Phi_n\rangle}\label{Res_For_Psi_+},\\
        \underset{\lambda=\lambda_n}{\operatorname{Res}} \Psi_-(u)
 &=\frac{2}{\pi}\frac{\Phi_n(u)}{ \langle\tilde{\Phi}_n| \widehat{K}\ket{\Phi_n}}
 \left[
 \left(1-\frac{\beta^2}{(1+\alpha)^2}\right)\braket{\Phi_n}{e_-}
 -\frac{i\pi^2\beta\lambda_n}{2(1+\alpha)^2}
       \braket{\Phi_n}{e_+}
 \right].\label{Res_For_Psi_-}
    \end{align}
\end{proposition}
\begin{proof}
At a spectral point $\lambda=\lambda_n$, the operator $\widehat L_{\lambda,\beta}$ becomes noninvertible. Since the eigenvalues are simple, its resolvent has at most a simple pole at each such point. The residue of an inhomogeneous solution is proportional to the corresponding eigenfunction and may vanish if the source has zero overlap with that eigenspace. Consider the inhomogeneous equation 
\begin{equation}
    \widehat{L}_{\lambda,\beta} \Psi_+=e_+.
\end{equation}
Its solution $\Psi_+$ is a meromorphic function in $\lambda$ with poles at the spectral points $\lambda_n$. Let us write the form of this function in the neighborhood of a pole as 
\begin{equation}\label{Psi_+-expansion}
    \Psi_+(u) = \frac{\Psi^{(-1)}_+(u)}{\lambda-\lambda_n}+\Psi^{(0)}_+(u)+\text{O}(\lambda-\lambda_n).
\end{equation}
Since $\Psi_+(u)$ is regular on the real axis, $\Psi^{(-1)}_+(u)$ and $\Psi^{(0)}_+(u)$ are also regular. It follows that 
\begin{equation}
    \widehat{L}_{\lambda,\beta}=\widehat{L}_{\lambda_n,\beta}-\frac{\pi}{2}(\lambda-\lambda_n) \widehat{K}.
\end{equation}
If we apply this operator to equation \eqref{Psi_+-expansion}, we get 
\begin{equation} \label{e+-labda-expansion}
    e_+(u)=\frac{\widehat{L}_{\lambda_n,\beta} \Psi^{(-1)}_+(u)}{\lambda-\lambda_n}+\widehat{L}_{\lambda_n,\beta} \Psi^{(0)}_+(u)-\frac{\pi}{2}\widehat{K}\Psi^{(-1)}_+(u)+\text{O}(\lambda-\lambda_n).
\end{equation}
The left-hand side of this equation has no poles in $\lambda$; therefore, since the spectrum is non-degenerate, it follows that 
\begin{equation}
   \widehat{L}_{\lambda_n,\beta} \Psi^{(-1)}_+(u)=0 \quad \implies \quad \Psi^{(-1)}_+(u) = c \Phi_n(u), \quad c \in \mathbb{R},
\end{equation}
where $\Phi_n(u)$ is a solution of the homogeneous equation $\widehat{L}_{\lambda_n,\beta} \Phi_n=0$. Setting $\lambda=\lambda_n$ and multiply \eqref{e+-labda-expansion}  on the left-hand side by $\bra{\tilde{\Phi}_n}$. Since $\Psi^{(0)}_+(u)$ is regular
\begin{equation}
    \braket{\tilde{\Phi}_n}{\widehat{L}_{\lambda_n,\beta} \Psi^{(0)}_+}=\braket{\widehat{L}^{\text{T}}_{\lambda_n,\beta}\tilde{\Phi}_n}{ \Psi^{(0)}_+}=0
\end{equation}
and we obtain 
\begin{equation}
    c=-\frac{2}{\pi}\frac{\langle\tilde{\Phi}_n|e_+\rangle}{\langle\tilde{\Phi}_n |\widehat{K} |\Phi_n\rangle}, \quad \langle\tilde{\Phi}_n|e_+\rangle=\braket{\Phi_n}{e_+}.
\end{equation}
Finally, 
\begin{equation}
    \underset{\lambda=\lambda_n}{\operatorname{Res}} \widehat{L}^{-1}_{\lambda,\beta} e_+= \underset{\lambda=\lambda_n}{\operatorname{Res}} \Psi_+(u)=-\frac{2 \Phi_n(u)}{\pi} \frac{\braket{\Phi_n}{e_+}}{\langle\tilde{\Phi}_n |\widehat{K} |\Phi_n\rangle}.
\end{equation}

In the case 
\begin{equation}
    \widehat{L}_{\lambda,\beta} \Psi_-=e_-
\end{equation}
the same argument cannot be applied directly because $\Psi_-(u)$ is singular at $u=0$ and
\begin{equation}
    \braket{\tilde{\Phi}_n}{\widehat{L}_{\lambda_n,\beta} \Psi^{(0)}_-} \neq 0.
\end{equation}
A term from the pole appears on the right-hand side. The easiest way to see this is by examining the equation
\begin{equation}
    \widehat{L}^{-1}_{\lambda,\beta} \widehat{L}_{\lambda_n,\beta} \left(M \Phi_n \right)=M \Phi_n+\frac{\pi}{2}(\lambda-\lambda_n)\widehat{L}^{-1}_{\lambda,\beta} \widehat{K }\left(M \Phi_n \right),
\end{equation}
from which it is immediately clear that $\widehat{L}^{-1}_{\lambda,\beta} \widehat{L}_{\lambda_n,\beta} \left(M \Phi_n \right)$ has no pole at the point $\lambda=\lambda_n$. Then we can apply the operator $\widehat{L}^{-1}_{\lambda,\beta}$ to \eqref{comutator_relation} and take the residue at $\lambda=\lambda_n$ from both sides of the equation. One obtains 
\begin{equation}
    \frac{\pi \lambda_n}{2} \braket{\Phi_n}{e_-} \underset{\lambda=\lambda_n}{\operatorname{Res}} \widehat{L}^{-1}_{\lambda,\beta} e_+=\left(\frac{\pi \lambda_n}{2}\braket{\Phi_n}{e_+}-\frac{i \beta}{\pi} \braket{\Phi_n}{e_-} \right) \underset{\lambda=\lambda_n}{\operatorname{Res}} \widehat{L}^{-1}_{\lambda,\beta} e_-.
\end{equation}
The commutator identity \eqref{comutator_relation}, together with
$\langle\widetilde\Phi_n|e_-\rangle=-\braket{\Phi_n}{e_-}$, gives
\begin{equation}\label{matrix-element-first}
 \begin{split}
 \langle\tilde\Phi_n|\widehat L_{\lambda_n,\beta}\ket{M\Phi_n}
 ={}&-\pi\lambda_n
       \braket{\Phi_n}{e_+}\braket{\Phi_n}{e_-}
       +\frac{i\beta}{\pi}\braket{\Phi_n}{e_-}^2.
 \end{split}
\end{equation}
Also using \eqref{For_overlaps-II} we obtain
\begin{equation}\label{matrix-element-second}
  \langle\tilde\Phi_n|\widehat L_{\lambda_n,\beta}\ket{M\Phi_n}
 =-\frac{i\pi\beta}{(1+\alpha)^2}
 \left(\frac{\pi\lambda_n}{2}\braket{\Phi_n}{e_+}
       -\frac{i\beta}{\pi}\braket{\Phi_n}{e_-}\right)^2.
\end{equation}
So we have 
\begin{equation}\label{matrix-element-comparison}
 \pi\lambda_n\braket{\Phi_n}{e_+}\braket{\Phi_n}{e_-}
 =\frac{i\beta}{\pi}\braket{\Phi_n}{e_-}^2+\frac{i\pi\beta}{(1+\alpha)^2}
 \left(\frac{\pi\lambda_n}{2}\braket{\Phi_n}{e_+}
       -\frac{i\beta}{\pi}\braket{\Phi_n}{e_-}\right)^2.
\end{equation}
The elementary identity
\begin{equation}
 \frac{i\beta}{\pi}\braket{\Phi_n}{e_-}^2
 =\frac{\pi\lambda_n}{2}
       \braket{\Phi_n}{e_+}\braket{\Phi_n}{e_-}-\left(\frac{\pi\lambda_n}{2}\braket{\Phi_n}{e_+}
       -\frac{i\beta}{\pi}\braket{\Phi_n}{e_-}\right)
       \braket{\Phi_n}{e_-}
\end{equation}
now factorizes \eqref{matrix-element-comparison} as
\begin{equation}\label{overlap-product-factorized}
 \begin{split}
 &\frac{\pi\lambda_n}{2}
       \braket{\Phi_n}{e_+}\braket{\Phi_n}{e_-}=\\
 &\quad=
 \left(\frac{\pi\lambda_n}{2}\braket{\Phi_n}{e_+}
       -\frac{i\beta}{\pi}\braket{\Phi_n}{e_-}\right)
 \left[
 -\braket{\Phi_n}{e_-}
 +\frac{i\pi\beta}{(1+\alpha)^2}
 \left(\frac{\pi\lambda_n}{2}\braket{\Phi_n}{e_+}
       -\frac{i\beta}{\pi}\braket{\Phi_n}{e_-}\right)
 \right].
 \end{split}
\end{equation}
Using this relation we can write the residue $\underset{\lambda=\lambda_n}{\operatorname{Res}} \widehat{L}^{-1}_{\lambda,\beta} e_-$ as
\begin{equation}
 \underset{\lambda=\lambda_n}{\operatorname{Res}}\widehat L^{-1}_{\lambda,\beta}e_-= \underset{\lambda=\lambda_n}{\operatorname{Res}} \Psi_-(u)
 =\frac{2}{\pi}\frac{\Phi_n(u)}{ \langle\tilde{\Phi}_n| \widehat{K}\ket{\Phi_n}}
 \left[
 \left(1-\frac{\beta^2}{(1+\alpha)^2}\right)\braket{\Phi_n}{e_-}
 -\frac{i\pi^2\beta\lambda_n}{2(1+\alpha)^2}
       \braket{\Phi_n}{e_+}
 \right].
\end{equation}
\end{proof}
\section{Spectral residues and the FLZ logarithmic-derivative identities}\label{log-relations}

In this section, we establish the pole structure and unit-residue properties underlying the logarithmic-derivative identities proposed in \cite{Artemev:2025cev}:
\begin{equation}\label{2.17--1}
\begin{aligned}
&\partial_\lambda \log D_-(\lambda)
-\frac{1}{\lambda}
\left(
    \frac{\beta^2}{2\alpha(\alpha+\beta)}
    -\frac{2i\beta G'(i)}{\pi^2}
\right)
-\frac{\pi^2\beta}{2\alpha(\alpha+\beta)}
-\frac{1}{4}(\alpha+\beta)\,\mathrm{i}_2(\alpha+\beta)
\\
&\qquad =
2i\left(1+\frac{\beta}{\pi^2\lambda}\right)
\left.
\partial_u \log \mathbf{Q}_+(u)
\right|_{u=i},
\end{aligned}
\end{equation}
and
\begin{equation}\label{2.17--2}
\begin{aligned}
&\partial_\lambda \log D_+(\lambda)
-\frac{1}{\lambda}
\left(
    \frac{\beta^2}{2\alpha}
    \frac{3\alpha+\beta}{\alpha^2-\beta^2}
    -\frac{2i\beta G'(i)}{\pi^2}
\right)
-\frac{\pi^2\beta}{2\alpha(\alpha+\beta)}
-\frac{1}{4}(\alpha+\beta)\,\mathrm{i}_2(\alpha+\beta)
\\
&\qquad =
2i\,
\frac{
    \left(1-\frac{\alpha}{\pi^2\lambda}\right)
    \mathbf{Q}'_-(i)
    +\frac{\alpha-\beta}{\pi^2\lambda}
    \mathbf{Q}'_-(-i)
}{
    \mathbf{Q}_-(i)
} =
2i\left(1-\frac{\alpha}{\pi^2\lambda}\right)
\left.
\partial_u \log \mathbf{Q}_-(u)
\right|_{u=i}
-
2i\frac{\alpha+\beta}{\pi^2\lambda}
\left.
\partial_u \log \mathbf{Q}_-(-u)
\right|_{u=i}.
\end{aligned}
\end{equation}

These equations contain the derivative of the function $G(u)$ defined in \cite{Artemev:2025cev}, $G'(i)=\frac{i}{8} ((\alpha+\beta)\mathrm{i}_2(\alpha+\beta) -(\alpha-\beta)\mathrm{i}_2(\alpha-\beta))$, where 
\begin{equation}
    \mathrm{i}_2(\alpha)= \int_{\mathbb{R}}\limits \frac{\sinh{t} (\sinh{2t} -2t)}{t \cosh^{2}{t} (\alpha \sinh{t}+t \cosh{t})} dt.
\end{equation}
In the case of general $\alpha$ and $\beta=0$ \eqref{2.17--1} and \eqref{2.17--2} simplify to:
\begin{equation}
    \partial_\lambda \log D_-(\lambda) - \frac{\alpha}{4} \mathrm{i}_2(\alpha)=2i \left.\partial_u \log Q_+(u) \right|_{u=i}, 
\end{equation}
\begin{equation}
    \partial_\lambda \log D_+(\lambda) - \frac{\alpha}{4} \mathrm{i}_2(\alpha)=2i \left(1-\frac{2 \alpha}{\pi^2} \lambda^{-1} \right)\left.\partial_u \log Q_-(u) \right|_{u=i}. 
\end{equation}

Differentiating \eqref{D_pm-def} we obtain
\begin{equation}
\begin{aligned}
    &\partial_\lambda \log D_-(\lambda)=\sum_{n=0}^\infty \left(\frac{1}{\lambda-\lambda_{2n+1}}+\frac{1}{n+1} \right)+\log 8\pi/e, \\ 
    &\partial_\lambda \log D_+(\lambda)=\sum_{n=0}^\infty \left(\frac{1}{\lambda-\lambda_{2n}}+\frac{1}{n+1} \right)+\log 8\pi/e.
\end{aligned}    
\end{equation}
To establish the agreement of the spectral principal parts of \eqref{2.17--1} and \eqref{2.17--2}, we must show that their right-hand sides have simple poles with unit residues at the eigenvalues of the corresponding spectral sectors:
\begin{equation}\label{Res=1}
    \underset{\lambda=\lambda_{2n+1}}{\operatorname{Res}} \mathcal{R}_-(\lambda)=\underset{\lambda=\lambda_{2n}}{\operatorname{Res}} \mathcal{R}_+(\lambda)=1.
\end{equation}
where 
\begin{equation}\label{Rpm}
    \mathcal{R}_-(\lambda)\overset{\text{def}}{=} 2i\left(1+\frac{\beta}{\pi^2\lambda}\right)
        \frac{\mathbf Q_+'(i|\lambda)}{\mathbf Q_+(i|\lambda)}, \quad \mathcal{R}_+(\lambda)\overset{\text{def}}{=}2i\frac{
       \left(1-\frac{\alpha}{\pi^2\lambda}\right)\mathbf Q_-'(i|\lambda)
       +\frac{\alpha-\beta}{\pi^2\lambda}\mathbf Q_-'(-i|\lambda)}
                         {\mathbf Q_-(i|\lambda)}.
\end{equation}
\subsection{Poles and zeros of the bold \texorpdfstring{$Q$}{Q}-functions}\label{Poles and zeros of the bold Q-functions}
To evaluate the residues of the FLZ expressions at the physical eigenvalues, we use the normalization constants $\mathcal N_n$ introduced in Section \ref{nondeg-section} and employed in the residue formulas of Section \ref{residue-section}. Recall that $\mathcal N_n$ was defined as a squared norm of the wave function in $x-$space by \eqref{N-norm}.  In  Fourier space, it converts to the bilinear pairing of the eigenfunction $\Phi_n$ with its reflected counterpart and the integral operator $\widehat K$
\begin{equation}\label{N-definition}
 \mathcal{N}_n= \langle\tilde\Phi_n|\widehat K\ket{\Phi_n},
 \qquad \tilde\Phi_n(u)=\Phi_n(-u).
\end{equation}

We now compare the residues of \eqref{2.17--1} and \eqref{2.17--2} at the physical spectral points. The bold combinations used below are
\begin{align}\label{residue-bold-combinations}
 \mathbf\Psi_+(u|\lambda)
 &=\Psi_+(u|\lambda)
   +\frac{2i\beta(1+\alpha)}{\pi^2\lambda(1+\alpha+\gamma)}
       \Psi_-(u|\lambda),\notag\\
 \mathbf\Psi_-(u|\lambda)
 &=\Psi_-(u|\lambda)
   -\frac{i\pi^2\beta\lambda}{2(1+\alpha)(1+\alpha+\gamma)}
       \Psi_+(u|\lambda).
\end{align}
Substituting \eqref{Res_For_Psi_+} and \eqref{Res_For_Psi_-} into
\eqref{residue-bold-combinations} and using $\beta^2=(1+\alpha)^2-\gamma^2$, we obtain the following expressions, valid in both spectral sectors:
\begin{align}
 \underset{\lambda=\lambda_n}{\operatorname{Res}}
       \mathbf\Psi_+(u|\lambda)
 &=\frac{2\gamma\Phi_n(u)}{\pi(1+\alpha)\mathcal{N}_n}
   \left[-\braket{\Phi_n}{e_+}
   +\frac{2i\beta\gamma}{\pi^2\lambda_n(1+\alpha+\gamma)}
        \braket{\Phi_n}{e_-}\right],
 \label{bold-Psi-residue-general-plus}\\
 \underset{\lambda=\lambda_n}{\operatorname{Res}}
       \mathbf\Psi_-(u|\lambda)
 &=\frac{2\gamma\Phi_n(u)}{\pi(1+\alpha)^2\mathcal{N}_n}
   \left[\gamma\braket{\Phi_n}{e_-}
   -\frac{i\pi^2\beta\lambda_n}{2(1+\alpha+\gamma)}
        \braket{\Phi_n}{e_+}\right].
 \label{bold-Psi-residue-general-minus}
\end{align}
The relations \eqref{overlaps-even-and-odd} now imply
\begin{align}
 \underset{\lambda=\lambda_{2n}}{\operatorname{Res}}
       \mathbf\Psi_+(u|\lambda)
 &=-\frac{4\gamma\Phi_{2n}(u)\braket{\Phi_{2n}}{e_+}}
          {\pi(1+\alpha+\gamma)\mathcal{N}_{2n}},
 &\underset{\lambda=\lambda_{2n}}{\operatorname{Res}}
       \mathbf\Psi_-(u|\lambda)&=0,
 \label{bold-Psi-residues-even}\\
 \underset{\lambda=\lambda_{2n+1}}{\operatorname{Res}}
       \mathbf\Psi_-(u|\lambda)
 &=\frac{4\gamma^2\Phi_{2n+1}(u)\braket{\Phi_{2n+1}}{e_-}}
          {\pi(1+\alpha)(1+\alpha+\gamma)\mathcal{N}_{2n+1}},
 &\underset{\lambda=\lambda_{2n+1}}{\operatorname{Res}}
       \mathbf\Psi_+(u|\lambda)&=0.
 \label{bold-Psi-residues-odd}
\end{align}
Multiplication by $\sinh \frac{\pi u}{2}$ gives the corresponding residues
of $\mathbf Q_\pm$. In particular, since $\sinh \frac{\pi i}{2}=i$,
\begin{align}
 \underset{\lambda=\lambda_{2n}}{\operatorname{Res}}
       \mathbf Q_+(i|\lambda)
 &=-\frac{4i\gamma\Phi_{2n}(i)\braket{\Phi_{2n}}{e_+}}
          {\pi(1+\alpha+\gamma)\mathcal{N}_{2n}},
 \label{bold-Q-pole-even}\\
 \underset{\lambda=\lambda_{2n+1}}{\operatorname{Res}}
       \mathbf Q_-(i|\lambda)
 &=\frac{4i\gamma^2\Phi_{2n+1}(i)\braket{\Phi_{2n+1}}{e_-}}
          {\pi(1+\alpha)(1+\alpha+\gamma)\mathcal{N}_{2n+1}}.
 \label{bold-Q-pole-odd}
\end{align}

Recall the factorized Wronskian obtained above:
\begin{equation}\label{residue-Wronskian-factorization}
 \mathbf Q_+(i|\lambda)\mathbf Q_-(i|\lambda)=\frac{W(0)}{P(\lambda)},
 \qquad
 P(\lambda)=\frac{\pi^2\lambda}{\pi^2\lambda+\beta}\,
       \frac{\alpha+\beta}{\alpha-\beta}\,
       \frac{1+\alpha+\gamma}{\gamma}.
\end{equation}
At a generic spectral point, its right-hand side is finite and nonzero. Thus, a simple pole of one factor is accompanied by a
simple zero of the other:
\begin{equation}\label{bold-Q-spectral-zeros}
 \mathbf Q_+(i|\lambda_{2n+1})=0,
 \qquad \mathbf Q_-(i|\lambda_{2n})=0.
\end{equation}
The same zeros occur at $u=-i$ by
\eqref{linear-relations-for-bold-Q}--\eqref{linear-relations-for-bold-Q-2}. These are statements about the special-point values; the regular $Q$-function at a spectral point does not vanish identically as a function of $u$.

Taking spectral residues in \eqref{linear-relations-for-bold-Q}--\eqref{linear-relations-for-bold-Q-2},
and using $\sinh(\frac{-i\pi}{2})=-i$, also yields:
\begin{align}
 \Phi_{2n+1}(-i)
 &=-\frac{\alpha+\beta}{\alpha-\beta}\Phi_{2n+1}(i),
 \label{Phi-ratio-odd}\\
 \Phi_{2n}(-i)
 &=\frac{\pi^2\lambda_{2n}-\beta}{\pi^2\lambda_{2n}+\beta}\,\,
     \frac{\alpha+\beta}{\alpha-\beta}\Phi_{2n}(i).
 \label{Phi-ratio-even}
\end{align}
\subsection{The regular \texorpdfstring{$Q$}{Q}-functions at spectral points}\label{Q-derivitive}
We next evaluate the derivatives entering the numerators of \eqref{2.17--1} and \eqref{2.17--2}. At spectral points, \eqref{e_+_bold},  \eqref{e_-_bold} and definitions of $\mathbf{\Psi}_\pm(u|\lambda)$ \eqref{bold-Psi-def} show that
\begin{align}
    M(u)\Phi_{2n+1}(u)&=-\frac{\pi \lambda_{2n+1}}{2} \braket{\Phi_{2n+1}}{e_-} \mathbf{\Psi}_+(u|\lambda_{2n+1}),\\
    M(u)\Phi_{2n}(u)&=\frac{\pi \lambda_{2n}}{2} \frac{1+\alpha}{\gamma} \braket{\Phi_{2n}}{e_+} \mathbf{\Psi}_-(u|\lambda_{2n}).
\end{align}
Multiplying these identities by
$\sinh\frac{\pi u}{2}$, we obtain
\begin{align}
 \mathbf Q_+(u|\lambda_{2n+1})
 &=-\frac{2}
          {\pi\lambda_{2n+1}\braket{\Phi_{2n+1}}{e_-}}\cosh\frac{\pi u}{2}\,\Phi_{2n+1}(u),
 \label{regular-bold-Q-odd}\\
 \mathbf Q_-(u|\lambda_{2n})
 &=\frac{2\gamma}
          {\pi\lambda_{2n}(1+\alpha)\braket{\Phi_{2n}}{e_+}}\cosh\frac{\pi u}{2}\,\Phi_{2n}(u).
 \label{regular-bold-Q-even}
\end{align}
Differentiating with respect to $u$ at $u=\pm i$, we find
\begin{align}
 \mathbf Q_+'(i|\lambda_{2n+1})
 &=-\frac{i\Phi_{2n+1}(i)}
          {\lambda_{2n+1}\braket{\Phi_{2n+1}}{e_-}},
 \label{bold-Q-prime-odd}\\
 \mathbf Q_-'(i|\lambda_{2n})
 =\frac{i\gamma\Phi_{2n}(i)}
          {\lambda_{2n}(1+\alpha)\braket{\Phi_{2n}}{e_+}}&,\quad
 \mathbf Q_-'(-i|\lambda_{2n})
 =-\frac{i\gamma\Phi_{2n}(-i)}
          {\lambda_{2n}(1+\alpha)\braket{\Phi_{2n}}{e_+}}.
 \label{bold-Q-prime-even}
\end{align}
\subsection{Evaluation of spectral residues}
Using the value \eqref{Wronskian-at-zero} gives
\begin{equation}\label{residue-P-over-W}
 \frac{P(\lambda)}{W(0)}
 =\frac{\pi^2\lambda}{\pi^2\lambda+\beta}
   \frac{(\alpha+\beta)^2}{\alpha^2}
   \frac{(1+\alpha)(1+\alpha+\gamma)}{2i\gamma^2}.
\end{equation}
We can therefore write the right-hand sides of
\eqref{2.17--1} and \eqref{2.17--2} without division by a $Q$-function:
\begin{align}
 &\mathcal{R}_-(\lambda)=\frac{(\alpha+\beta)^2(1+\alpha)(1+\alpha+\gamma)}
        {\alpha^2\gamma^2}
       \mathbf Q_+'(i|\lambda)\mathbf Q_-(i|\lambda),
 \label{log-RHS-odd-product}\\
 \mathcal{R}_+(\lambda)=&
 \frac{(\alpha+\beta)^2(1+\alpha)(1+\alpha+\gamma)}
      {\alpha^2\gamma^2(\pi^2\lambda+\beta)}
 \left[(\pi^2\lambda-\alpha)\mathbf Q_-'(i|\lambda)
                 +(\alpha-\beta)\mathbf Q_-'(-i|\lambda)\right]
       \mathbf Q_+(i|\lambda).
 \label{log-RHS-even-product}
\end{align}

At $\lambda=\lambda_{2n+1}$ $\mathbf Q_+'(i|\lambda)$ in \eqref{log-RHS-odd-product} is regular, while $\mathbf Q_-(i|\lambda)$ has
the pole \eqref{bold-Q-pole-odd}. The complete substitution is
\begin{equation}\label{log-residue-odd-before-N}
 \begin{aligned}
 \underset{\lambda=\lambda_{2n+1}}{\operatorname{Res}}
 \mathcal{R}_-(\lambda)=
 \frac{(\alpha+\beta)^2(1+\alpha)(1+\alpha+\gamma)}
      {\alpha^2\gamma^2}
 \left(-\frac{i\Phi_{2n+1}(i)}
              {\lambda_{2n+1}\braket{\Phi_{2n+1}}{e_-}}\right)
 \frac{4i\gamma^2\Phi_{2n+1}(i)\braket{\Phi_{2n+1}}{e_-}}
      {\pi(1+\alpha)(1+\alpha+\gamma)\mathcal{N}_{2n+1}}=\\
 =\frac{4(\alpha+\beta)^2\Phi_{2n+1}(i)^2}
                  {\pi\alpha^2\lambda_{2n+1}\mathcal{N}_{2n+1}}.
 \end{aligned}
\end{equation}

At $\lambda=\lambda_{2n}$ $\mathbf Q_-'(i|\lambda)$ and $\mathbf Q_-'(-i|\lambda)$ in
\eqref{log-RHS-even-product} are regular. Substituting both lines
of \eqref{bold-Q-prime-even} and the pole \eqref{bold-Q-pole-even}
gives
\begin{multline}\label{log-residue-even-before-N}
\underset{\lambda=\lambda_{2n}}{\operatorname{Res}}
 \mathcal{R}_+(\lambda)=\\=
 \frac{(\alpha+\beta)^2(1+\alpha)(1+\alpha+\gamma)}
              {\alpha^2\gamma^2(\pi^2\lambda_{2n}+\beta)}
 \left[(\pi^2\lambda_{2n}-\alpha)
       \frac{i\gamma\Phi_{2n}(i)}
       {\lambda_{2n}(1+\alpha)\braket{\Phi_{2n}}{e_+}}
       -(\alpha-\beta)
       \frac{i\gamma\Phi_{2n}(-i)}
       {\lambda_{2n}(1+\alpha)\braket{\Phi_{2n}}{e_+}}\right]\cdot\\
 \cdot\left[-\frac{4i\gamma\Phi_{2n}(i)\braket{\Phi_{2n}}{e_+}}
              {\pi(1+\alpha+\gamma)\mathcal{N}_{2n}}\right]=
 \frac{4(\alpha+\beta)^2}{\pi\alpha^2\lambda_{2n}\mathcal{N}_{2n}}
       \frac{\Phi_{2n}(i)
       \left[(\pi^2\lambda_{2n}-\alpha)\Phi_{2n}(i)
                   -(\alpha-\beta)\Phi_{2n}(-i)\right]}
            {\pi^2\lambda_{2n}+\beta}.
\end{multline}
All the remaining terms displayed on the left-hand sides of
\eqref{2.17--1} and \eqref{2.17--2} are holomorphic at these spectral points. In the opposite sector, each right-hand side is regular: in its quotient form the numerator has at most the same pole order as the denominator. To finish the comparison
it remains to evaluate $\mathcal{N}_n$ independently. Appendix \ref{N-evaluation} gives
precisely the values needed in \eqref{log-residue-odd-before-N} and
\eqref{log-residue-even-before-N}, directly from the homogeneous
equation and its analytic continuation:
\begin{equation}\label{N-general}
 \frac{\pi\lambda_n}{2}\mathcal{N}_n=
       \frac{(\alpha+\beta)^2}{\alpha^2}\Phi_n(i)^2
       +\frac{(\alpha-\beta)^2}{\alpha^2}\Phi_n(-i)^2-\frac{\left[(\alpha+\beta)\Phi_n(i)
                   +(\alpha-\beta)\Phi_n(-i)\right]^2}
             {\pi^2\alpha\lambda_n}.
\end{equation}

For odd spectral points, \eqref{Phi-ratio-odd} makes the square in the last line vanish and the first two terms equal. For even points, we use \eqref{Phi-ratio-even}. The result is
\begin{align}
 \mathcal{N}_{2n+1}
 &=\frac{4(\alpha+\beta)^2}{\pi\alpha^2\lambda_{2n+1}}
       \Phi_{2n+1}(i)^2,
 \label{N-odd}\\
 \mathcal{N}_{2n}
 &=\frac{4(\alpha+\beta)^2}{\pi\alpha^2\lambda_{2n}}
   \frac{\pi^4\lambda_{2n}^2-2\alpha\pi^2\lambda_{2n}+\beta^2}
        {(\pi^2\lambda_{2n}+\beta)^2}\Phi_{2n}(i)^2.
 \label{N-even}
\end{align}
Equivalently, the even normalization is
\begin{equation}\label{N-even-mixed}
 \mathcal{N}_{2n}
 =\frac{4(\alpha+\beta)^2}{\pi\alpha^2\lambda_{2n}}
 \frac{\Phi_{2n}(i)
       \left[(\pi^2\lambda_{2n}-\alpha)\Phi_{2n}(i)
                    -(\alpha-\beta)\Phi_{2n}(-i)\right]}
            {\pi^2\lambda_{2n}+\beta}.
\end{equation}
Substitution of \eqref{N-odd} and \eqref{N-even-mixed} into
\eqref{log-residue-odd-before-N} and
\eqref{log-residue-even-before-N}, respectively, gives a residue
equal to one in both cases, as required by
\eqref{Res=1}.
\subsection{Additional singularities and the remaining analytic ambiguity}
Equations \eqref{Wronskian-at-zero} and \eqref{Wronskian-at-zero-2} give the exact identity
\begin{equation}\label{Wronskian-factorized}
\mathbf{Q}_+(i|\lambda)\mathbf{Q}_-(i|\lambda)
=
\frac{2i\gamma^2\alpha^2}
{(1+\alpha)(1+\alpha+\gamma)(\alpha+\beta)^2}
\frac{\pi^2\lambda+\beta}{\pi^2\lambda},
\end{equation}
Now suppose $\lambda_*\neq0,-\beta/\pi^2$ is not an eigenvalue. Both functions on the left of \eqref{Wronskian-factorized} are regular there, whereas their product is finite and nonzero. Therefore,
\begin{equation}
\mathbf Q_+(i|\lambda_*)\neq0,
\qquad
\mathbf Q_-(i|\lambda_*)\neq0.
\end{equation}
This proves that the quotients defining $\mathcal{R}_-(\lambda)$ and $\mathcal{R}_+(\lambda)$ in \eqref{Rpm} have no additional poles at generic non-spectral points, except possibly $0$ and $-\beta/\pi^2$. The latter comes explicitly from the right-hand side of \eqref{Wronskian-factorized}. It can be shown that the residue at this apparent pole is exactly zero.

Thus, $\mathcal{R}_{\pm}(\lambda)$ are non-singular  at $\lambda=-\beta/\pi^2$ and have an apparent simple pole at $\lambda=0$. The explicit  subtraction terms in \eqref{2.17--1} and \eqref{2.17--2} aim to cancel this pole. Up to an additive constant, their exact form can be extracted from the explicit solution \cite{Artemev:2025cev} of the inhomogeneous Fredholm equation \eqref{inhomogeneous-equation}   for $\lambda=0$.

The preceding arguments show that the left- and right-hand sides of \eqref{2.17--1} and \eqref{2.17--2} have identical singularities in the finite complex $\lambda$-plane. In particular, their differences are entire functions of $\lambda$. To establish the spectral-determinant identities \eqref{2.17--1}--\eqref{2.17--2} globally, it remains to determine these entire functions. So far, we have not been able to determine them rigorously, but the numerical analysis performed in \cite{Artemev:2025cev} strongly supports their disappearance.
\section{FLZ integrability}\label{FLZ-integrability}
The results obtained in this paper suggest a possible extension of the FLZ approach to a broader class of integral equations with completely integrable kernels. An important question is which properties of the 't Hooft kernel are essential for the existence of relations between spectral determinants and special-point values of Baxter $Q$-functions. Although the general theory of integrable Fredholm operators provides a natural framework for addressing this question, the derivation presented here relies on additional algebraic and analytic properties that are not shared by arbitrary integrable kernels.

Consider an integral operator with the kernel
\begin{equation}\label{generic-kernel}
K(u,v)=
\frac{h(u)h(v)\bigl(g(u)-g(v)\bigr)}
{M(u)-M(v)}\frac{1}{f(v)},
\qquad
f(u)=c+g(u)M(u),
\end{equation}
where $M(u)$, $g(u)$ and $h(u)$ are meromorphic functions satisfying suitable regularity and asymptotic conditions. The 't Hooft kernel belongs to this class, with
\begin{equation}
g(u)=u,\qquad
M(u)=\coth\frac{\pi u}{2},\qquad
h(u)=i\sqrt{\frac{\pi}{2}}\frac{1}{\sinh(\pi u/2)},
\qquad c=\frac{2\alpha}{\pi}.
\end{equation}
We also introduce the companion operator
\begin{equation}
S(u,v)=\frac{h(u)h(v)}{M(u)-M(v)}
\frac{1}{f(v)}.
\end{equation}

With respect to the bilinear pairing
\begin{equation}
\langle a|b\rangle
=\int_{\mathbb R}\frac{a(u)b(u)}{f(u)}du,
\end{equation}
the operators satisfy
\begin{equation}\label{generic-comm-relat}
[\widehat{M},\widehat{K}]=|gh\rangle\langle h|-|h\rangle\langle gh|,
\qquad
[\widehat{M},\widehat{S}]=|h\rangle\langle h|,
\end{equation}
where $\widehat{M}$ denotes the $M(u)$ multiplication operator. Furthermore,
\begin{equation}
\widehat{K}^T=\widehat{K},\qquad \widehat{S}^T=-\widehat{S}.
\end{equation}

Consequently, the operator family
\begin{equation}
\widehat{L}_{\lambda,\beta}=1-\beta \widehat{S}-\lambda \widehat{K}
\end{equation}
obeys the transpose relation $\widehat{L}_{\lambda,\beta}^T=\widehat{L}_{\lambda,-\beta}$. These identities are independent of the particular functional form of $M$, $g$ and $h$. They show that the finite-rank commutator structure underlying our analysis is shared by a broad family of rank-two integrable kernels.

A second important ingredient is the relation $f=c+gM$. Indeed, introducing $H(u)=M(u)^{-1}$, one obtains the elementary identity
\begin{equation}\label{kernel-expansion}
\frac{g(u)-g(v)}{f(u)f(v)}=\frac{H(u)}{f(v)}
-\frac{H(v)}{f(u)}
-c\frac{H(u)-H(v)}{f(u)f(v)}.
\end{equation}

Since
\begin{equation}
\frac{H(u)-H(v)}{M(u)-M(v)}
=-H(u)H(v),
\end{equation}
the last contribution becomes a rank-one separable kernel. This is a generalization of the algebraic decomposition used in our evaluation of the eigenfunction normalization factors. It suggests that at least part of the normalization calculation can be extended to the entire family under consideration.

We now restrict ourselves to $\beta=0$ and describe a sufficient set of hypotheses under which this mechanism gives explicit logarithmic-derivative formulas.   A natural possibility is to assume the existence of a complex shift $T$ such that
\begin{equation}\label{generic-shift}
M(u+T)=M(u),\qquad
g(u+T)=g(u)+\delta,\qquad
h(u+T)=\sigma h(u),
\end{equation}
where $\delta$ and $\sigma$ are constants and $\sigma^2=1$. In the 't Hooft model, $T=2i$, $\delta=2i$ and $\sigma=-1$. Under these assumptions, the kernels obey
\begin{equation}
S(u+T,v)=\sigma S(u,v),
\end{equation}
\begin{equation}
K(u+T,v)=
\sigma\bigl[K(u,v)+\delta S(u,v)\bigr].
\end{equation}

We impose the following analytic assumptions
\begin{itemize}
    \item The functions $M,g,h,f$ must admit meromorphic continuation to the strips used in the contour deformations. On the integration contour, the original kernel must be regular after its removable singularities are cancelled.
    \item $\widehat{K}$ defines a compact Fredholm problem with isolated, simple spectral points. Its transpose is taken with respect to the bilinear pairing already specified.
    \item In one period strip, $M$ has one simple zero and one simple
pole, represented by $a$ and $p$, respectively:
\begin{equation}
M(a)=0,\qquad M'(a)\ne0,\qquad
M(u)=\frac{\underset{z=p}{\operatorname{Res}}M(z)}{u-p}+O(1).
\end{equation}
We choose $0<\operatorname{Im}a\le\operatorname{Im}T$ and
$0\le\operatorname{Im}p<\operatorname{Im}T$.
Thus, a zero on the real axis is represented by its translate on the upper boundary. The function $h$ has simple poles at
$p+kT$, no zeros in the strips used below, and no other poles
there. The function $g$ is holomorphic in these strips.
The moving singularities of the kernel come only from
$u-v=kT$, $k \in \mathbb{Z}$; the singularity at $u=v$ is removable.
In particular, no other preimages of $M(v)$ are crossed in
the continuation by $T$. The weight $f=c+gM$ is regular and strictly positive on
$\mathbb R$ after removable singularities have been filled in.
If $p$ is real, we require $g(p)=0$, so the pole of $M$ cancels
in $f$.
\item
The solutions
\begin{equation}\label{Fredholm}
(1-\lambda_n \widehat{K})\Phi_n=0,\qquad
(1-\lambda \widehat{K})\Psi_+=gh,\qquad
(1-\lambda \widehat{K})\Psi_-=h
\end{equation}
admit the continuation
\begin{equation}
Q_n(u)=\frac{\Phi_n(u)}{h(u)},\qquad
Q_\pm(u,\lambda)=\frac{\Psi_\pm(u,\lambda)}{h(u)}.
\label{flz:Q}
\end{equation}
These $Q$-functions are holomorphic in the closed strip
$|\operatorname{Im}u|\le\operatorname{Im}T$ and grow
subexponentially in $\operatorname{Re}u$, uniformly in the
strip.
\item All operator identities and contour deformations below are
assumed valid in the admissible classes. The homogeneous
eigenfunctions are regular at the relevant common poles of
$M$ and $h$, and all contour integrals converge. In particular, this condition
applies to $h^2(u)Q_n(u)Q_n(u-T)/M(u)$.
\end{itemize}

Continuing the integral equation through its moving poles gives
\begin{equation}\label{TQ-flz}
Q(u+T)+Q(u-T)-2Q(u) = 2\pi i\lambda\delta\,
\frac{h(u)^2}{M'(u)f(u)}Q(u).
\end{equation}
The same equation holds for $Q_n$ and $Q_\pm$.

Define
\begin{equation}
\mathcal N_n=\langle\Phi_n|\widehat{K}|\Phi_n\rangle,\qquad
\langle\Phi_n|\Phi_n\rangle=\lambda_n\mathcal N_n.
\end{equation}
Under the assumptions above and \eqref{kernel-expansion}, the normalization is
\begin{equation}\label{N-explicit}
\mathcal N_n= \frac{c}{\lambda_n^2\delta^2}
\bigl[Q_n(a)-Q_n(a-T)\bigr]^2
-\frac{2\pi i\,h(a)^2}{\lambda_n\delta M'(a)}
Q_n(a)Q_n(a-T).
\end{equation}
This expression is obtained through calculations similar to those shown in Appendix \ref{N-evaluation}.

In the absence of diagonal poles at a simple spectral point the residues have the universal form
\begin{equation}\label{Psi-residues}
\operatorname*{Res}_{\lambda=\lambda_n}\Psi_\pm(u,\lambda)
=
-\frac{\Phi_n(u)}{\mathcal N_n}
\langle\Phi_n|e_\pm\rangle,
\qquad e_+=gh,\quad e_-=h,
\end{equation}
where
\begin{equation}
\langle\Phi_n|e_+\rangle
=\fint_{\mathbb R}\frac{g(u)h(u)\Phi_n(u)}{f(u)}du,
\qquad
\langle\Phi_n|e_-\rangle
=\fint_{\mathbb R}\frac{h(u)\Phi_n(u)}{f(u)}du.
\end{equation}
They can also be obtained from local data at a pole of $M$. Indeed, expanding homogeneous equation at points $p$ and $p-T$ one finds 
\begin{align}
\langle\Phi_n|h\rangle
&=
\frac{\underset{z=p}{\operatorname{Res}}M(z)}{\lambda_n\delta}
\bigl[Q_n'(p)-Q_n'(p-T)\bigr],\\
\langle\Phi_n|gh\rangle
&=
\frac{\underset{z=p}{\operatorname{Res}}M(z)}{\lambda_n\delta}
\left([g(p)-\delta]Q_n'(p)-g(p)Q_n'(p-T)\right)
\end{align}
for $0<\operatorname{Im}p<\operatorname{Im}T$. And for real $p$ such that $g(p)=0$ the boundary prescription
instead gives
\begin{equation}
\langle\Phi_n|gh\rangle
=-\frac{\underset{z=p}{\operatorname{Res}}M(z)}{\lambda_n}Q_n'(p), \quad
\langle\Phi_n|h\rangle
=
\frac{\underset{z=p}{\operatorname{Res}}M(z)}{\lambda_n\delta}
\bigl[Q_n'(p+T)-Q_n'(p)\bigr]
+\frac{i\pi\bigl(\underset{z=p}{\operatorname{Res}}h(z)\bigr)^2}{f(p)}
Q_n'(p).
\end{equation}
The last term follows from the half-residue in the upper
continuation formula. It must be retained even though the
homogeneous eigenfunction itself is regular at $p$. 

In other words, the residues  of the functions $\Psi_\pm$ at points in the spectrum can be expressed in terms of the values of $Q_n$ and $Q_n'$ at a finite number of points.

The $TQ$ equation implies that
\begin{equation}
W(u,\lambda)=
Q_+(u,\lambda)Q_-(u-T,\lambda)
-Q_-(u,\lambda)Q_+(u-T,\lambda)
\end{equation}
is $T$-periodic. Analyticity in one period strip and
subexponential growth make it independent of $u$.
For an interior pole $p$, the source normalization gives
$Q_+(p)=g(p)$, $Q_+(p-T)=g(p)-\delta$,
and $Q_-(p)=Q_-(p-T)=1$.
For a real pole, it gives $Q_+(p)=0$,
$Q_+(p+T)=\delta$, and $Q_-(p)=1$.
Consequently, in both configurations,
\begin{equation}
W(u,\lambda)=\delta.
\end{equation}

An additional property is needed to turn these local formulas into scalar logarithmic-derivative relations. We require that the special-point values satisfy
\begin{equation}\label{Q-rel-for-flz}
Q_\pm(a-T,\lambda)=d_\pm(\lambda)Q_\pm(a,\lambda),
\end{equation}
where $d_\pm$ are explicitly known meromorphic functions
obtained independently from the equation and its normalization.
They are not defined as ratios of the unknown solutions.
This is a condition at two specified points, not a parity
condition on the entire functions.
It must be verified in each proposed example. At a simple spectral point, suppose that $d_\pm(\lambda)$ are regular
and $d_+(\lambda_n)\ne d_-(\lambda_n)$. Taking residues of
\eqref{Q-rel-for-flz} and using \eqref{Psi-residues} gives
\begin{equation}
\langle\Phi_n|gh\rangle\,
\langle\Phi_n|h\rangle=0,
\end{equation}
provided $Q_n(a)\ne0$.  Thus, exactly one source overlap is nonzero, separating the
spectrum into two disjoint sectors: $\Sigma_+$ couples to
$gh$, while $\Sigma_-$ couples to $h$. Here  $\Sigma_+=\{\lambda_n: \langle\Phi_n|gh\rangle \neq 0 \}$ and $\Sigma_-=\{\lambda_n: \langle\Phi_n|h\rangle \neq 0 \}$. Correspondingly,
$Q_+$ has spectral poles only in $\Sigma_+$, and $Q_-$ only
in $\Sigma_-$. These sector labels need not coincide with
the parity of the spectral index $n$ \footnote{Similarly, in the two-flavor 't Hooft model, separation into spectral sectors is defined in \eqref{For_overlaps-I} and \eqref{For_overlaps-II}. 

In the case $\beta=0$ we have $\mathcal{B}_{2n+1}=0$ because $\Phi_{2n+1}(u)$ is odd and $\mathcal{A}^{(2n+1)}_+=\mathcal{A}^{(2n+1)}_-$. \eqref{For_overlaps-II} gives $\mathcal{A}_{2n}=0.$
}.

The Wronskian now gives the explicit product identity
\begin{equation}
Q_+(a,\lambda)Q_-(a,\lambda)
=\frac{\delta}{d_-(\lambda)-d_+(\lambda)}.
\label{flz:product}
\end{equation}
At a spectral point where the right-hand side is finite and
nonzero, a pole of one factor is paired with a zero of the
other.

Arguing as in Section \ref{Q-derivitive} from commutation relation 
\begin{equation}
(1-\lambda_n \widehat{K})(M\Phi_n) = \lambda_n\left( gh\langle\Phi_n|h\rangle-h\langle\Phi_n|gh\rangle
\right).
\end{equation}
one can express the values of the derivatives of the $Q$-functions at a point in the spectrum. The final answer for the residue of the logarithmic derivative is 
\begin{equation}
\operatorname*{Res}_{\lambda=\lambda_n}
\partial_u\log Q_\mp(u,\lambda)|_{u=a} =
\pm\frac{M'(a)
[d_-(\lambda_n)-d_+(\lambda_n)]Q^2_n(a)}
{\lambda_n\delta \mathcal N_n}.
\end{equation}
An explicit expression for $\mathcal{N}_n$ is given in \eqref{N-explicit}. Consider 
\begin{equation}\label{generic-R}
\mathcal R_\pm(\lambda)=
\frac{\pm1}{M'(a)[d_-(\lambda)-d_+(\lambda)]}
\left[
\frac{c[1-d_\pm(\lambda)]^2}{\lambda\delta}
-\frac{2\pi i h^2(a)d_\pm(\lambda)}{M'(a)}
\right] 
\partial_u\log Q_\mp(u,\lambda)|_{u=a}.
\end{equation}
The functions $\mathcal R_\pm$ have residue one at the
spectral points in $\Sigma_\pm$ and are regular at those
in $\Sigma_\mp$. If $\widehat K$ is Hilbert--Schmidt, define
$D_\pm$ as regularized spectral products over $\Sigma_\pm$,
up to exponential normalization. Then $\mathcal R_\pm$ and
$\partial_\lambda\log D_\pm$ have the same spectral
principal parts. This comparison fixes only the spectral principal parts.
Possible nonspectral singularities must be analysed independently,
for example using local perturbative expansions of the inhomogeneous Fredholm equation at the exceptional values of $\lambda$. Establishing a global identity also requires control
of the remaining entire ambiguity, which may be achieved through suitable growth bounds and independently established large-$\lambda$ asymptotics. 

Note that similar expressions (Eq. (4.15) from \cite{Litvinov:2025geb}), involving a log-derivative in the Ising field theory, are obtained when 
\begin{align}
    M(u)&=\tanh{\frac{\pi u}{2}}, \quad h(u)=\sqrt{\frac{\pi}{2}}\frac{1}{\cosh{\frac{\pi u}{2}}},  \quad g(u)=u, \\ c&=\frac{2 \alpha}{\pi},  \quad p=i, \quad a=2i, \quad T=\delta=2i, \quad d_+(\lambda)=0,  \quad d_-(\lambda)= \left(1-\frac{\pi^2
 \lambda}{\alpha}\right)^{-1}.
\end{align}
The constant in Eq. (4.15) depends on the normalization of the spectral determinants and can be determined using an explicit perturbative solution derived in the same paper.

A closely related approach to FLZ integrability has been developed independently in \cite{BercelBoldisa2026}. In Section 5 of that work, the authors consider a broader class of spectral problems with the same integral kernels as in the QCD and Ising models. Under certain algebraic conditions, they derive relations between logarithmic derivatives of spectral determinants and values of the corresponding Baxter $Q$-functions at finitely many distinguished points. Section 6 of \cite{BercelBoldisa2026} extends this approach to more general integrable kernels using finite-rank commutator identities. Our construction starts from the kernel \eqref{generic-kernel}, with $f=c+gM$, and relies on the commutation relations \eqref{generic-comm-relat}, the shift properties \eqref{generic-shift}, and the special-point relations \eqref{Q-rel-for-flz}. These ingredients allow us to express the normalization factors of eigenfunctions through special-point values of the Baxter functions and establish the unit-residue property of the expressions \eqref{generic-R}. Thus, both approaches exploit the same underlying finite-rank structure of integrable Fredholm operators but impose different conditions on the kernels and their associated Baxter functions. The results of \cite{BercelBoldisa2026} are mathematically more complete in this respect since they establish global spectral-determinant identities for the classes of operators considered there. Our construction, on the other hand, establishes the equality of spectral principal parts, while the cancellation of possible non-spectral singularities and the determination of the remaining entire-function ambiguity require further analysis.

The mechanism described above consists of a finite-rank commutator, a scalar $TQ$ equation selected by analyticity, an explicit reduction of eigenfunction norms to special-point data, and independently known relations such as \eqref{Q-rel-for-flz}. We propose  to use FLZ integrability for spectral problems in which these ingredients close and reconstruct the spectral determinants from finitely many Baxter data, allowing explicit finite-rank extensions.

The calculation above establishes the equality of the spectral residues. To conclude an identity for the entire meromorphic functions, one must also match non-spectral singularities and fix the remaining entire contribution by global growth bounds
and independently controlled large-$\lambda$ asymptotics.
\section{Concluding remarks}
We have investigated the Fredholm and Baxter formulations of the ’t Hooft equation for arbitrary unequal quark masses. Using finite-rank commutator identities, we derived relations between the canonical Baxter solutions at distinguished complex points and identified the combinations associated with the two spectral sectors. We also established nondegeneracy of the meson spectrum and calculated the residues of the inhomogeneous solutions at the physical eigenvalues. Expressing the eigenfunction normalization factors through their values at $u=\pm i$, we proved that the proposed FLZ expressions have the unit spectral residues required by the logarithmic derivatives of the corresponding spectral determinants.

These results clarify the analytic mechanism underlying the FLZ identities, although their complete proof still requires control of the remaining entire-function ambiguity. We also identified structural ingredients that may permit similar constructions for a broader class of integrable integral operators. Establishing global analytic estimates and extending the method to other spectral problems are natural directions for further investigation.
\section*{Note added}
During the final stages of preparing this paper, we learned that most of the results had been independently derived in \cite{BercelBoldisa2026}. We thank the authors for sharing their manuscript with us and for coordinating the submission of our papers to arXiv.
\section*{Acknowledgments}
We acknowledge discussions with Alexander Artemev, Marco Bertola, Pavlo Gavrylenko, Oleg Lisovyy, Pavel Meshcheriakov, Vitaly Tarasov and Alexander Zamolodchikov. A.L. thanks the organizers of the workshop “Algebraic Geometry and Integrable Systems,” held at TSIMF in December 2025, for creating a stimulating scientific atmosphere, and all participants for fruitful discussions. The work of A.L. was supported by the Russian Science Foundation under the grant 26-11-00342. The work of A.L. and E.S. was supported by the BASIS Foundation.
The work of N.S. was performed within the framework of the state assignment at Steklov
Mathematical Institute of Russian Academy of Sciences.
\Appendix
\section{Useful integral identities}\label{integrals}
Let us take the homogeneous equation
\begin{equation}\label{Fredholm_homogeneous-appendix}
     \Phi_n(u) - \frac{i\beta}{\pi} \fint_{\mathbb{R}}\limits \frac{dv}{f(v)}\frac{1}{\sinh{\frac{\pi (u-v)}{2}}} \Phi_n(v) - \frac{\pi\lambda_n}{2} \int_{\mathbb{R}}\limits \frac{dv}{f(v)} \frac{ \pi (u-v)}{\sinh{\frac{\pi (u-v)}{2}}} \Phi_n(v)=0,
\end{equation}
multiply by
\begin{equation}
\frac{\coth \frac{\pi u}{2}\Phi_n(-u)}{f(u)},
\end{equation}
and take the principal value integral. 

The first integral vanishes
\begin{equation}
\fint_{\mathbb{R}}\limits\coth \frac{\pi u}{2}\Phi_n(-u)\Phi_n(u)\frac{du}{f(u)}=0,
\end{equation}
since this is an integral of an odd function. 

For the second integral, we represent
\begin{equation}
    4\fint_{\mathbb{R}}\limits\fint_{\mathbb{R}}\limits dudv
    =
    \underbrace{\int_{\mathbb{R}+2i\epsilon}\limits dv
    \int_{\mathbb{R}+i\epsilon}\limits du}_{\mathrm{I}}+
    \underbrace{\int_{\mathbb{R}+2i\epsilon}\limits du
    \int_{\mathbb{R}+i\epsilon}\limits dv}_{\mathrm{II}}+
    \underbrace{\int_{\mathbb{R}-i\epsilon}\limits du
    \int_{\mathbb{R}-2i\epsilon}\limits dv}_{\mathrm{III}}+
    \underbrace{\int_{\mathbb{R}-i\epsilon}\limits dv
    \int_{\mathbb{R}-2i\epsilon}\limits du}_{\mathrm{IV}}.
\end{equation}
Take $\mathrm{III}$ and replace $u\to -v$, $v\to -u$, then we have
\begin{equation}
    \mathrm{I}+\mathrm{III}=
    -\int_{\mathbb{R}+2i\epsilon}\limits dv
    \int_{\mathbb{R}+i\epsilon}\limits du\,
    \Phi_n(-u)\Phi_n(v)
    \frac{1}{\sinh\frac{\pi u}{2}\sinh\frac{\pi v}{2}}
    \frac{1}{f(u)f(v)}
    \overset{\eqref{eq:A-B}}{=}\mathcal{A}_+^{(n)}\mathcal{A}_-^{(n)}.
\end{equation}
The integrals $\mathrm{II}$ and $\mathrm{IV}$ combine in the same way. Thus, we have
\begin{equation}
    \fint_{\mathbb{R}}\limits\coth \frac{\pi u}{2}\Phi_n(-u)\frac{du}{f(u)}\fint_{\mathbb{R}}\limits \frac{dv}{f(v)}\frac{1}{\sinh{\frac{\pi (u-v)}{2}}} \Phi_n(v)=
    \frac{1}{2}\mathcal{A}_+^{(n)}\mathcal{A}_-^{(n)}.
\end{equation}

Similar calculation shows
\begin{equation}
    \fint_{\mathbb{R}}\limits\coth \frac{\pi u}{2}\Phi_n(-u)\frac{du}{f(u)}\int_{\mathbb{R}}\limits \frac{dv}{f(v)} \frac{(u-v)}{\sinh{\frac{\pi (u-v)}{2}}} \Phi_n(v)=
    -\frac{1}{2}\mathcal{B}_n
    \left(\mathcal{A}_+^{(n)}+\mathcal{A}_-^{(n)}\right).
\end{equation}
\section{Relations between \texorpdfstring{$A_\pm$ and $B_\pm$}{A± and B±}}\label{details}
Here we present the detailed calculations necessary to prove Proposition \ref{linear-relations-prop}. 

By \eqref{singular-source-normalization} and $f(0)=2(1+\alpha)/\pi$,
\begin{equation}\label{ratio-origin-residue}
 \frac{\Psi_-(u)}{f(u)\cosh\frac{\pi u}2}
 =\frac{1}{(1+\alpha)u}+O(1),\qquad
 \underset{u=0}{\operatorname{Res}}
 \frac{\Psi_-(u)}{f(u)\cosh\frac{\pi u}2}
 =\frac{1}{1+\alpha}.
\end{equation}
The lower integral minus the upper integral of $\Psi_-(u)/[f(u)\cosh\frac{\pi u}2]$ is therefore
$2\pi i/(1+\alpha)$. Combining this difference with
\eqref{contour_def} gives
\begin{equation}\label{A-contour-values}
 \begin{aligned}
 \int_{\mathbb{R}+i \epsilon}\limits\frac{\Psi_-(u)}{f(u)\cosh\frac{\pi u}2}du
 =&A_--\frac{i\pi}{1+\alpha},\quad
 \int_{\mathbb{R}-i\epsilon}\limits\frac{\Psi_-(u)}{f(u)\cosh\frac{\pi u}2}du
 =A_-+\frac{i\pi}{1+\alpha},\\
 \int_{\mathbb{R}+i\epsilon}\limits\frac{\Psi_+(u)}{f(u)\cosh\frac{\pi u}2}du
 &=\int_{\mathbb{R}-i\epsilon}\limits\frac{\Psi_+(u)}{f(u)\cosh\frac{\pi u}2}du
 =A_+.
 \end{aligned}
\end{equation}
The last equality uses the regularity of $\Psi_+$ at zero. The
integrands defining $B_+$ and $B_-$ are also regular there: in $B_-$
the explicit factor $u$ cancels the pole. Their upper- and
lower-contour values consequently coincide. Our goal is to multiply \eqref{ratio-inhomogeneous-plus}-\eqref{ratio-inhomogeneous-minus} by
$\tanh\frac{\pi u}{2}\Psi_+(-u)/f(u)$ and by $\tanh\frac{\pi u}{2}\Psi_-(-u)/f(u)$, respectively, and take the contour half-sum. This will allow us to express $A_\pm$ and $B_\pm$ in terms of each other.

\subsection{One identity from each inhomogeneous equation}
First consider the singular solution $\Psi_-$. Take the $u$ contour
above the real axis and the $v$ contour below it. All contour
deformations are initially made with nonzero heights, so the pole
at $u=v$ never lies on either contour. The change of variables
$(u,v)\mapsto(-v,-u)$ preserves this product of contours and gives
\begin{equation}\label{ratio-opposite-reflection}
 \begin{aligned}
 \int_{\mathbb{R}+i \epsilon}\limits\!du\int_{\mathbb{R}-i\epsilon}\limits\!dv\,
 \frac{\tanh\frac{\pi u}2\,\Psi_-(-u)\Psi_-(v)}
 {f(u)f(v)\sinh\frac{\pi(u-v)}2}
 =-\int_{\mathbb{R}+i\epsilon}\limits\!du\int_{\mathbb{R}-i \epsilon}\limits\!dv\,
 \frac{\tanh\frac{\pi v}2\,\Psi_-(-u)\Psi_-(v)}
 {f(u)f(v)\sinh\frac{\pi(u-v)}2}.
 \end{aligned}
\end{equation}
We used $f(-u)=f(u)$ and the oddness of $\tanh \frac{\pi u}{2}$.
Averaging the two expressions and using
\begin{equation}\label{ratio-hyperbolic-identity}
 \frac{\tanh\frac{\pi u}2-\tanh\frac{\pi v}2}
 {\sinh\frac{\pi(u-v)}2}
 =\frac{1}{\cosh\frac{\pi u}2\cosh\frac{\pi v}2}
\end{equation}
separates the two integrations:
\begin{equation}\label{ratio-separated-product}
 \begin{aligned}
 \int_{\mathbb{R}+i \epsilon}\limits\!du\int_{\mathbb{R}-i \epsilon}\limits\!dv\,
 \frac{\tanh\frac{\pi u}2\,\Psi_-(-u)\Psi_-(v)}
 {f(u)f(v)\sinh\frac{\pi(u-v)}2}=\frac12
 \left(\,\int_{\mathbb{R}+i\epsilon}\limits
 \frac{\Psi_-(-u)}{f(u)\cosh\frac{\pi u}2}du\,\right)
 \left(\,\int_{\mathbb{R}-i \epsilon}\limits
 \frac{\Psi_-(v)}{f(v)\cosh\frac{\pi v}2}dv\,\right).
 \end{aligned}
\end{equation}
Substituting the variables yields 
\begin{equation}\label{ratio-reflected-single-integral}
 \int_{\mathbb{R}+i\epsilon}\limits\frac{\Psi_-(-u)}{f(u)\cosh\frac{\pi u}2}du
 =\int_{\mathbb{R}-i \epsilon}\limits\frac{\Psi_-(v)}{f(v)\cosh\frac{\pi v}2}dv
 =A_-+\frac{i\pi}{1+\alpha}.
\end{equation}
The two factors in \eqref{ratio-separated-product} are thus values
on the same lower contour. Interchanging the upper and lower contours
gives the analogous square on the upper contour:
\begin{equation}\label{T-opposite-contours}
 \begin{aligned}
 &\int_{\mathbb{R}+i\epsilon}\limits\!du\int_{\mathbb{R}-i\epsilon}\limits\!dv\,
 \frac{\tanh\frac{\pi u}2\,\Psi_-(-u)\Psi_-(v)}
 {f(u)f(v)\sinh\frac{\pi(u-v)}2}
 =\frac12\left(A_-+\frac{i\pi}{1+\alpha}\right)^2,\\
 &\int_{\mathbb{R}-i\epsilon}\limits\!du\int_{\mathbb{R}+i \epsilon}\limits\!dv\,
 \frac{\tanh\frac{\pi u}2\,\Psi_-(-u)\Psi_-(v)}
 {f(u)f(v)\sinh\frac{\pi(u-v)}2}
 =\frac12\left(A_--\frac{i\pi}{1+\alpha}\right)^2.
 \end{aligned}
\end{equation}

The double integrals in \eqref{T-opposite-contours} are
related to the meromorphic convolution $\widehat S\Psi_-$ that occurs
in the original equation. 

We begin by multiplying \eqref{ratio-inhomogeneous-minus} by $\tanh{\frac{\pi u}{2}} \Psi_-(-u)/f(u)$ and integrating it in the sense of \eqref{contour_def}. The local product is odd, so
\begin{equation}\label{ratio-plus-local-source}
 \begin{aligned}
 \fint_{\mathbb{R}}\limits
 \frac{\tanh\frac{\pi u}2\,\Psi_-(-u)\Psi_-(u)}{f(u)}du=0, \quad
  \fint_{\mathbb{R}}\limits
 \frac{\tanh\frac{\pi u}2\,\Psi_-(-u)}{f(u)}
 \frac{1}{\sinh\frac{\pi u}2}\,du
 =\fint_{\mathbb{R}}\limits\frac{\Psi_-(-u)}{f(u)\cosh\frac{\pi u}2}\,du=A_-.
 \end{aligned}
\end{equation}
Consider the action of such integral operator on $\widehat{S} \Psi_-$. There are two relevant residues in its
integration variable:
\begin{equation}\label{ratio-two-residues}
 \underset{v=u}{\operatorname{Res}}
 \frac{\Psi_-(v)}{f(v)\sinh\frac{\pi(u-v)}2}
 =-\frac{2\Psi_-(u)}{\pi f(u)}, \quad 
 \underset{v=0}{\operatorname{Res}}
 \frac{\Psi_-(v)}{f(v)\sinh\frac{\pi(u-v)}2}
 =\frac{1}{(1+\alpha)\sinh\frac{\pi u}2}.
\end{equation}
To separate the two operations, put $u$ initially at height
$\varepsilon>0$ and the two inner contours at heights
$\delta$ and $-\delta$, with $0<\delta<\varepsilon$. The half-sum
of these inner integrals has the source pole averaged at $v=0$.
Its difference from the meromorphic continuation from real $u$ is
$-2i\Psi_-(u)/f(u)$, the half-residue of the diagonal pole in
\eqref{ratio-two-residues}. The lower inner integral exceeds its
half-sum by $i\pi/[(1+\alpha)\sinh(\pi u/2)]$. For negative
$\operatorname{Im}u$, both signs are reversed. Consequently,
\begin{equation}\label{T-inner-contour-transfer}
 \begin{aligned}
 \int_{\mathbb{R}-i \epsilon}\limits
 \frac{\Psi_-(v)}{f(v)\sinh\frac{\pi(u-v)}2}dv
 &=(\widehat S\Psi_-)(u)-\frac{2i\Psi_-(u)}{f(u)}
   +\frac{i\pi}{(1+\alpha)\sinh\frac{\pi u}2},
 &&\operatorname{Im}u>0,\\
 \int_{\mathbb{R}+i \epsilon}\limits
 \frac{\Psi_-(v)}{f(v)\sinh\frac{\pi(u-v)}2}dv
 &=(\widehat S\Psi_-)(u)+\frac{2i\Psi_-(u)}{f(u)}
   -\frac{i\pi}{(1+\alpha)\sinh\frac{\pi u}2},
 &&\operatorname{Im}u<0.
 \end{aligned}
\end{equation}
These equations are used in a sufficiently small strip about the
real axis. No other singularity is crossed. We now multiply the first and the second equations in
\eqref{T-inner-contour-transfer} by
$\tanh \frac{\pi u}{2}\Psi_-(-u)/f(u)$ and integrate over the upper and lower contour respectively. Then we take their half-sum. 
The resulting three terms can be evaluated separately. Using \eqref{singular-source-normalization} it is easy to see that 
\begin{equation}\label{res-II}
     -2i \int_{\mathbb{R}+i \epsilon}\limits\tanh\frac{\pi u}{2} \cdot
 \frac{\Psi_-(-u)\Psi_-(u)}{f^2(u)}\,du
 +2i \int_{\mathbb{R}-i \epsilon}\limits \tanh\frac{\pi u}{2} \cdot
 \frac{\Psi_-(-u)\Psi_-(u)}{f^2(u)}\,du=-\frac{2 \pi^2}{(1+\alpha)^2}.
\end{equation}
The third addends in the right hand side of \eqref{T-inner-contour-transfer} after summation give
\begin{equation}\label{res-III}
     \frac{i \pi}{1+\alpha} \int_{\mathbb{R}+i \epsilon}\limits 
 \frac{\Psi_-(-u)}{\cosh{\frac{\pi u}{2}}}\,\frac{du}{f(u)}
 -\frac{i \pi}{1+\alpha} \int_{\mathbb{R}-i \epsilon}\limits \frac{\Psi_-(-u)}{\cosh{\frac{\pi u}{2}}}\,\frac{du}{f(u)}=\frac{2 \pi^2}{(1+\alpha)^2}.
\end{equation}
We find that the terms \eqref{res-II} and \eqref{res-III} cancel each other out. So only left hand side's contribution remains. Using \eqref{T-opposite-contours} one obtains
\begin{equation}\label{Psi-S-Psi_-}
 \fint_{\mathbb{R}}\limits
 \frac{\tanh\frac{\pi u}2\,\Psi_-(-u)}{f(u)}
 (\widehat S\Psi_-)(u)\,du
 =\frac14\left[
 \left(A_-+\frac{i\pi}{1+\alpha}\right)^2
 +\left(A_--\frac{i\pi}{1+\alpha}\right)^2\right]
 =\frac12\left(A_-^2-\frac{\pi^2}{(1+\alpha)^2}\right).
\end{equation}

For $\Psi_+$, there is no source pole in
\eqref{ratio-two-residues}. The diagonal correction is an integral
of a regular odd function and vanishes. Repeating the same steps
therefore gives
\begin{equation}\label{Psi-S-Psi_+}
 \fint_{\mathbb{R}}\limits
 \frac{\tanh\frac{\pi u}2\,\Psi_+(-u)}{f(u)}
 (\widehat S\Psi_+)(u)\,du=\frac12 A_+^2.
\end{equation}
This regular-case calculation uses only analyticity at the origin,
decay, and reflection of the contours.

For $\widehat K$, the factor $u-v$ removes the diagonal pole. One may
symmetrize the iterated half-sum directly. For the singular solution,
\begin{equation}\label{Psi-K-Psi_-}
 \begin{aligned}
 &\fint_{\mathbb{R}}\limits
 \frac{\tanh\frac{\pi u}2\,\Psi_-(-u)}{f(u)}
 (\widehat K\Psi_-)(u)\,du
 =\frac12\fint_{\mathbb{R}}\limits\!du\,\fint_{\mathbb{R}}\limits\!dv\,
 \frac{(u-v)\Psi_-(-u)\Psi_-(v)}
 {f(u)f(v)\cosh\frac{\pi u}2\cosh\frac{\pi v}2}=\\
 &\quad=\frac12\left[
 \left(\int_{\mathbb{R}}\limits\frac{u\Psi_-(-u)}{f(u)\cosh\frac{\pi u}2}du\right)
 \left(\fint_{\mathbb{R}}\limits\frac{\Psi_-(v)}{f(v)\cosh\frac{\pi v}2}dv\right)
-
 \left(\fint_{\mathbb{R}}\limits\frac{\Psi_-(-u)}{f(u)\cosh\frac{\pi u}2}du\right)
 \left(\int_{\mathbb{R}}\limits\frac{v\Psi_-(v)}{f(v)\cosh\frac{\pi v}2}dv\right)
 \right]=\\
 &\quad=\frac12\bigl[(-B_-)A_--A_-B_-\bigr]=-A_-B_-.
 \end{aligned}
\end{equation}
Reflection changes the sign of the moment with $u$, while it
preserves the contour half-sum of the moment without $u$.
The same computation for $\Psi_+$ gives
\begin{equation}\label{Psi-K-Psi_+}
 \fint_{\mathbb{R}}\limits
 \frac{\tanh\frac{\pi u}2\,\Psi_+(-u)}{f(u)}
 (\widehat K\Psi_+)(u)\,du=-A_+B_+.
\end{equation}

Integrating  \eqref{ratio-inhomogeneous-plus}-\eqref{ratio-inhomogeneous-minus} with
$\tanh\frac{\pi u}{2}\Psi_+(-u)/f(u)$ and with $\tanh\frac{\pi u}{2}\Psi_-(-u)/f(u)$, respectively, then combining equations \eqref{Psi-S-Psi_-} -- \eqref{Psi-K-Psi_+}  we obtain
\begin{align}
    &-\frac{i\beta}{2\pi}A_+^2
 +\frac{\pi\lambda}{2}A_+B_+=-B_+,\\
  &-\frac{i\beta}{2\pi}
 \left(A_-^2-\frac{\pi^2}{(1+\alpha)^2}\right)
 +\frac{\pi\lambda}{2}A_-B_-=A_-. 
\end{align}

\subsection{The mixed identity from the two equations}
Multiply the equation with source $e_+$ by
$\tanh \frac{\pi u}{2}\Psi_-(-u)/f(u)$, and the equation with source $e_-$
by $\tanh\frac{\pi u}{2}\Psi_+(-u)/f(u)$. We add the two equations before
integrating. Their local contribution is
\begin{equation}\label{mixed-local-integral}
 \fint_{\mathbb{R}}\limits\frac{\tanh\frac{\pi u}2}{f(u)}
 \bigl[\Psi_+(u)\Psi_-(-u)+\Psi_+(-u)\Psi_-(u)\bigr]\,du=0.
\end{equation}
The expression in square brackets is even.

For the singular convolution, repeat the opposite-contour
symmetrization with the sum of the two cross products. The factor
$1/2$ from symmetrization is cancelled by the two equal separated
products. Explicitly,
\begin{equation}\label{ratio-mixed-opposite-products1}
 \begin{aligned}
 &\int_{\mathbb{R}+i \epsilon}\limits\!du\int_{\mathbb{R}-i \epsilon}\limits\!dv\,
 \frac{\tanh\frac{\pi u}2
 [\Psi_+(-u)\Psi_-(v)+\Psi_-(-u)\Psi_+(v)]}
 {f(u)f(v)\sinh\frac{\pi(u-v)}2}=\\
 &\qquad=
 \left(\; \int_{\mathbb{R}-i\epsilon}\limits\frac{\Psi_+(u)}{f(u)\cosh\frac{\pi u}2}du\; \right)
 \left(\; \int_{\mathbb{R}-i\epsilon}\limits\frac{\Psi_-(v)}{f(v)\cosh\frac{\pi v}2}dv \; \right)
 =A_+\left(A_-+\frac{i\pi}{1+\alpha}\right),
 \end{aligned}
\end{equation}
\begin{equation}\label{ratio-mixed-opposite-products2}
 \begin{aligned}
 &\int_{\mathbb{R}-i\epsilon}\limits\!du\int_{\mathbb{R}+i\epsilon}\limits\!dv\,
 \frac{\tanh\frac{\pi u}2
 [\Psi_+(-u)\Psi_-(v)+\Psi_-(-u)\Psi_+(v)]}
 {f(u)f(v)\sinh\frac{\pi(u-v)}2}=\\
 &\qquad=
 \left(\;\int_{\mathbb{R}+i \epsilon}\limits\frac{\Psi_+(u)}{f(u)\cosh\frac{\pi u}2}du\;\right)
 \left(\;\int_{\mathbb{R}+i \epsilon}\limits\frac{\Psi_-(v)}{f(v)\cosh\frac{\pi v}2}dv\right)
 =A_+\left(A_--\frac{i\pi}{1+\alpha}\;\right).
 \end{aligned}
\end{equation}
From relations similar to \eqref{T-inner-contour-transfer}, it is easy to obtain 
\begin{equation}\label{T-S-mixed}
 \fint_{\mathbb{R}}\limits\frac{\tanh\frac{\pi u}2}{f(u)}
 \bigl[\Psi_+(-u)(\widehat S\Psi_-)(u)
       +\Psi_-(-u)(\widehat S\Psi_+)(u)\bigr]\,du=A_+A_-.
\end{equation}
For the regular convolution, the same separation as in
\eqref{Psi-K-Psi_-} gives
\begin{equation}\label{T-K-mixed}
 \begin{aligned}
 &\fint_{\mathbb{R}}\limits\frac{\tanh\frac{\pi u}2}{f(u)}
 \bigl[\Psi_+(-u)(\widehat K\Psi_-)(u)
       +\Psi_-(-u)(\widehat K\Psi_+)(u)\bigr]\,du\\
 &\qquad=\frac12\bigl[-B_+A_--A_+B_--B_-A_+-A_-B_+\bigr]
 =-(A_+B_-+A_-B_+).
 \end{aligned}
\end{equation}
Finally, the two source terms integrate to
\begin{equation}\label{ratio-mixed-source}
 \begin{aligned}
 &\fint_{\mathbb{R}}\limits\frac{\tanh\frac{\pi u}2}{f(u)}
 \left[\frac{u\Psi_-(-u)}{\sinh\frac{\pi u}2}
       +\frac{\Psi_+(-u)}{\sinh\frac{\pi u}2}\right]du=\\
 &\qquad=\int_{\mathbb{R}}\limits
 \frac{u\Psi_-(-u)}{f(u)\cosh\frac{\pi u}2}\,du
 +\fint_{\mathbb{R}}\limits\frac{\Psi_+(-u)}{f(u)\cosh\frac{\pi u}2}\,du
 =-B_-+A_+.
 \end{aligned}
\end{equation}
Substituting \eqref{mixed-local-integral}, \eqref{T-S-mixed},
\eqref{T-K-mixed}, and \eqref{ratio-mixed-source} gives
\begin{equation}
     -\frac{i\beta}{\pi}A_+A_-
 +\frac{\pi\lambda}{2}(A_+B_-+A_-B_+)=A_+-B_-.
\end{equation}
All three identities \eqref{A-B-plus}, \eqref{A-B-minus}, and
\eqref{A-B-mixed} have now been obtained from the two original
inhomogeneous equations, with their contour prescriptions retained.

\section{Contour evaluation of normalization constant \texorpdfstring{$\mathcal{N}_n$}{N}}\label{N-evaluation}
Here we evaluate the normalization constant (see \eqref{N-norm} and \eqref{N-definition})
\begin{equation}\label{normalization-matrix-element}
 \mathcal{N}_n=\langle{\tilde\Phi_n}|\widehat K\ket{\Phi_n},
 \qquad \tilde\Phi_n(u)=\Phi_n(-u),
\end{equation}
directly from the homogeneous equation
\eqref{Fredholm_homogeneous2}. The scalar product is bilinear, as
defined above, and no normalization of $\Phi_n$ is imposed. All singular integrals are understood as the contour half-sum \eqref{contour_def}, with both contours oriented from left to right.

Using \eqref{kernel-def}, we write
\begin{equation}\label{N-double-integral}
 \mathcal{N}_n
 =\int_{\mathbb R}\limits du\int_{\mathbb R}\limits dv\,
 \frac{(u-v)\Phi_n(-u)\Phi_n(v)}
      {f(u)f(v)\sinh\frac{\pi(u-v)}2}.
\end{equation}
The kernel in this expression is regular at $u=v$. In particular,
the ordinary integrals in \eqref{N-double-integral} agree with the
contour prescription.

The useful identity, which follows from the explicit form of the function $f(u)$, is
\begin{equation}\label{normalization-kernel-identity}
 \frac{u-v}{f(u)f(v)}
 =\frac{\tanh\frac{\pi u}2}{f(v)}
  -\frac{\tanh\frac{\pi v}2}{f(u)}
  -\frac{2\alpha}{\pi}
    \frac{\sinh\frac{\pi(u-v)}2}
    {f(u)f(v)\cosh\frac{\pi u}2\cosh\frac{\pi v}2}.
\end{equation}
Substituting it into \eqref{normalization-matrix-element}, we get
\begin{equation}\label{normalization-three-integrals}
 \begin{aligned}
 \frac{\pi\lambda_n}{2}\mathcal{N}_n
 ={}&\frac{\pi\lambda_n}{2}
 \int_{\mathbb R}\limits du\,\Phi_n(-u)
 \fint_{\mathbb R}\limits dv\,\Phi_n(v)
 \frac{1}{\sinh\frac{\pi(u-v)}2}
 \left[\frac{\tanh\frac{\pi u}2}{f(v)}
       -\frac{\tanh\frac{\pi v}2}{f(u)}\right]-\\
 &-\alpha\lambda_n
 \left(\int_{\mathbb R}\limits
          \frac{\Phi_n(-u)}{f(u)\cosh\frac{\pi u}2}du\right)
 \left(\int_{\mathbb R}\limits
          \frac{\Phi_n(v)}{f(v)\cosh\frac{\pi v}2}dv\right).
 \end{aligned}
\end{equation}
The integrand in the first line has no pole at $u=v$. When its two
terms are separated, each is understood with the contour half-sum
\eqref{contour_def}. In the second term in square brackets, set
$u=-v'$ and $v=-u'$. Evenness of $f$ and oddness of $\tanh$ show
that, including its minus sign, this term equals the first one.
Reflection interchanges the two contour prescriptions and therefore
preserves their half-sum. Similarly, the two integrals in the last
line coincide. Thus
\begin{equation}\label{N-reduction}
 \frac{\pi\lambda_n}{2}\mathcal{N}_n
 =\pi\lambda_n\int_{\mathbb R}\limits
       du\,\tanh\frac{\pi u}2\,\Phi_n(-u)
       \fint_{\mathbb R}\limits dv
       \frac{\Phi_n(v)}{f(v)\sinh\frac{\pi(u-v)}2}-\alpha\lambda_n
       \left(\int_{\mathbb R}\limits
          \frac{\Phi_n(u)}{f(u)\cosh\frac{\pi u}2}du\right)^2.
\end{equation}

To calculate these two terms, we use the homogeneous equation at
$u=i$, $u=-i$, and at the shifted arguments $u\pm2i$.
The first two substitutions are the homogeneous versions of
\eqref{inhomogeneous-at-i}:
\begin{align}
 \frac{\alpha+\beta}{\alpha}\Phi_n(i)
 ={}&\left(\frac{\beta}{\pi}+\frac{\pi\lambda_n}{2}\right)
       \int_{\mathbb R}\limits
       \frac{\Phi_n(v)}{f(v)\cosh\frac{\pi v}2}dv
 +\frac{i\pi\lambda_n}{2}
       \int_{\mathbb R}\limits
       \frac{v\Phi_n(v)}{f(v)\cosh\frac{\pi v}2}dv,\notag\\
 \frac{\alpha-\beta}{\alpha}\Phi_n(-i)
 ={}&\left(-\frac{\beta}{\pi}+\frac{\pi\lambda_n}{2}\right)
       \int_{\mathbb R}\limits
       \frac{\Phi_n(v)}{f(v)\cosh\frac{\pi v}2}dv
 -\frac{i\pi\lambda_n}{2}
       \int_{\mathbb R}\limits
       \frac{v\Phi_n(v)}{f(v)\cosh\frac{\pi v}2}dv.
 \label{normalization-at-i}
\end{align}
Adding them gives
\begin{equation}\label{normalization-A}
 \int_{\mathbb R}\limits
       \frac{\Phi_n(v)}{f(v)\cosh\frac{\pi v}2}dv
 =\frac{(\alpha+\beta)\Phi_n(i)+(\alpha-\beta)\Phi_n(-i)}
             {\pi\alpha\lambda_n}.
\end{equation}
Therefore, the second term of \eqref{N-reduction} is already known:
\begin{equation}\label{normalization-separated-term}
 -\alpha\lambda_n
 \left(\int_{\mathbb R}\limits
       \frac{\Phi_n(v)}{f(v)\cosh\frac{\pi v}2}dv\right)^2
 =-\frac{\left[(\alpha+\beta)\Phi_n(i)
                    +(\alpha-\beta)\Phi_n(-i)\right]^2}
              {\pi^2\alpha\lambda_n}.
\end{equation}

Let us now consider the first term. When continuing the homogeneous
equation away from the real axis, one must keep the diagonal
residue of the singular convolution. With the integration contour
fixed on $\mathbb R$,
\begin{equation}\label{normalization-boundary-values}
 \int_{\mathbb R\pm i0}\limits
       \frac{\Phi_n(v)}
            {f(v)\sinh\frac{\pi(u-v)}2}dv
 =(\widehat S\Phi_n)(u)\mp\frac{2i\Phi_n(u)}{f(u)}.
\end{equation}
Consequently, the continued equation is
\begin{equation}\label{homogeneous-off-axis}
 \left(1\pm\frac{2\beta}{\pi f(z)}\right)\Phi_n(z)
 =\frac{i\beta}{\pi}\int_{\mathbb R}\limits
       \frac{\Phi_n(v)}{f(v)\sinh\frac{\pi(z-v)}2}dv+\frac{\pi\lambda_n}{2}\int_{\mathbb R}\limits
       \frac{(z-v)\Phi_n(v)}{f(v)\sinh\frac{\pi(z-v)}2}dv,
 \qquad 0<\pm\operatorname{Im}z<2.
\end{equation}
Using \eqref{normalization-boundary-values} and \eqref{homogeneous-off-axis} one can obtain
\begin{align}
 &\left(1+\frac{2\beta}{\pi f(u+2i)}\right)\Phi_n(u+2i)
 -\left(1-\frac{2\beta}{\pi f(u-2i)}\right)\Phi_n(u-2i)
 =\frac{4\beta\Phi_n(u)}{\pi f(u)}
       -2\pi i\lambda_n(\widehat S\Phi_n)(u),
 \label{shifted-Phi-difference}\\
 &\left(1+\frac{2\beta}{\pi f(u+2i)}\right)\Phi_n(u+2i)
 +\left(1-\frac{2\beta}{\pi f(u-2i)}\right)\Phi_n(u-2i)
 =\left(\frac{4\pi\lambda_n}{f(u)}-2\right)\Phi_n(u).
 \label{homogeneous-TQ-Phi}
\end{align}
The second equality is also \eqref{TQ-equation} written for
$\Phi_n$. Substitute the first one into \eqref{N-reduction}.
The term containing $\beta\tanh(\frac{\pi u}{2})\Phi_n(-u)\Phi_n(u)/f(u)$
integrates to zero because it is odd. Hence
\begin{equation}\label{N-through-shifted-integral}
 \begin{aligned}
 \frac{\pi\lambda_n}{2}\mathcal{N}_n
 ={}&\frac{i}{2}\int_{\mathbb R}\limits
 du\,\tanh\frac{\pi u}2\,\Phi_n(-u)
 \left[\left(1+\frac{2\beta}{\pi f(u+2i)}\right)\Phi_n(u+2i)
      -\left(1-\frac{2\beta}{\pi f(u-2i)}\right)\Phi_n(u-2i)\right]-\\
 &-\frac{\left[(\alpha+\beta)\Phi_n(i)
                   +(\alpha-\beta)\Phi_n(-i)\right]^2}
             {\pi^2\alpha\lambda_n}.
 \end{aligned}
\end{equation}
It remains to evaluate the difference of the two shifted integrals by moving their contours. Also, due to the oddness of the right-hand side term in \eqref{homogeneous-TQ-Phi}, we have 
\begin{align*}
    \int_{\mathbb R}\limits
 du\,\tanh\frac{\pi u}2\,\Phi_n(-u)
 \left(1+\frac{2\beta}{\pi f(u+2i)}\right)\Phi_n(u+2i)=\\
= - \int_{\mathbb R}\limits
 du\,\tanh\frac{\pi u}2\,\Phi_n(-u)
 \left(1-\frac{2\beta}{\pi f(u-2i)}\right)\Phi_n(u-2i).
\end{align*}
If we define
\begin{equation}
    \mathcal{I}_\pm=\int_{\mathbb R}\limits
 du\,\tanh\frac{\pi u}2\,\Phi_n(-u)
 \left(1\pm\frac{2\beta}{\pi f(u\pm 2i)}\right)\Phi_n(u\pm 2i),
\end{equation}
then
\begin{equation}
 \frac{\pi\lambda_n}{2}\mathcal{N}_n
 =\frac{i}{2}\left(\mathcal{I_+}-\mathcal{I_-} \right)-
 \frac{\left[(\alpha+\beta)\Phi_n(i)
                   +(\alpha-\beta)\Phi_n(-i)\right]^2}
             {\pi^2\alpha\lambda_n}, \quad \mathcal{I_+}=-\mathcal{I_-}.
\end{equation}
So now our goal is to compute $\mathcal{I}_\pm$. For that let us consider functions 
\begin{equation}
    h_\pm(u)=\left(1\pm\frac{2 \beta}{\pi f(u)} \right) \Phi_n(u).
\end{equation}
As stated in \eqref{poles-of-Q}, $\Phi_n$ may have poles in the interval $-2<\text{Im }u<2$; however, for functions $h_\pm(u)$, these poles are canceled out by the expression in parentheses, which means that $h_+(u)$ is holomorphic in $0<\text{Im }u<2$ and $h_-(u)$ is holomorphic in $-2<\text{Im }u<0$ . The $TQ$ equation \eqref{homogeneous-TQ-Phi} can also be rewritten in terms of the functions $h_\pm(u)$:
\begin{equation}\label{TQ-for-h}
    h_+(u+2i)+h_-(u-2i)=\left(-2+\frac{4 \pi \lambda_n}{f(u)} \right) \Phi_n(u).
\end{equation}
Now we define 
\begin{equation}
    \mathcal{F}(u)=\frac{1}{2} \tanh{\frac{\pi u}{2}} [h_+(u)h_+(2i-u)+h_-(u-2i)h_-(-u)].
\end{equation}
By the stated analyticity properties of the functions $h_\pm(u)$ in the intervals described above, $\mathcal{F}(u)$ has only one pole in $u=i$ due to $\tanh \frac{\pi u}{2}$ in the interval $0<\text{Im }u<2$. It also satisfies the relationship
\begin{equation}
    \mathcal{F}(2i-u)=-\mathcal{F}(u).
\end{equation}

Consider
\begin{equation}
    \int_{\mathbb{R}+2i}\limits \mathcal{F}(u)du=\frac{1}{2}\int_{\mathbb{R}}\limits \tanh \frac{\pi u}{2}[h_+(u+2i)h_+(-u)+h_-(u) h_-(-u-2i)]du.
\end{equation}
After substituting $u \rightarrow -u$ in the second term, we get
\begin{equation}
    \int_{\mathbb{R}+2i}\limits \mathcal{F}(u)du=\frac{\mathcal{I}_+-\mathcal{I}_-}{2}+\frac{\beta}{\pi} \int_{\mathbb{R}}\limits \frac{\tanh \frac{\pi u}{2}}{f(u)} \Phi_n(-u) [h_+(u+2i)+h_-(u-2i)]du.
\end{equation}
The second term here is equal to zero because of \eqref{TQ-for-h} and 
\begin{equation} \label{I+-1}
    \mathcal{I}_+=-\mathcal{I}_-= \int_{\mathbb{R}+2i}\limits \mathcal{F}(u)du.
\end{equation}
Also, a sequence of simple variable substitutions yields
\begin{equation}\label{I+-2}
    \mathcal{I}_+=\int_{\mathbb{R}}\limits \mathcal{F}(u+2i)du=\int_{\mathbb{R}}\limits \mathcal{F}(2i-u)du=-\int_{\mathbb{R}}\limits \mathcal{F}(u)du.
\end{equation}
Combining \eqref{I+-1} and \eqref{I+-2}
\begin{equation}\label{shifted-integral-value}
    2 \mathcal{I}_+=\int_{\mathbb{R}+2i}\limits \mathcal{F}(u)du-\int_{\mathbb{R}}\limits \mathcal{F}(u)du=-2 \pi i \underset{u=i}{\operatorname{Res}} \mathcal{F}(u)=-2i [h^2_+(i)+h^2_-(-i)].
\end{equation}
Substituting \eqref{shifted-integral-value} into
\eqref{N-through-shifted-integral} gives
\begin{equation}
 \frac{\pi\lambda_n}{2}\mathcal{N}_n=
       \frac{(\alpha+\beta)^2}{\alpha^2}\Phi_n(i)^2
       +\frac{(\alpha-\beta)^2}{\alpha^2}\Phi_n(-i)^2-\frac{\left[(\alpha+\beta)\Phi_n(i)
                   +(\alpha-\beta)\Phi_n(-i)\right]^2}
             {\pi^2\alpha\lambda_n}.
\end{equation}

Finally, using the definition \eqref{a-b-definition} and the relations \eqref{A--B-homogeneous}-\eqref{homogeneous-at-i} one finds
\begin{equation}
    \mathcal{N}_n=\left(\pi\lambda_n-\frac{2\alpha}{\pi}\right)a_n^2-\pi\lambda_n b_n^2.
\end{equation}
\bibliographystyle{MyStyle.bst}
\bibliography{references.bib}

\end{document}